\documentclass[VANCOUVER,STIX1COL, doublespace]{WileyNJD-v2}
\usepackage{moreverb}
\usepackage[latin1]{inputenc}

\usepackage{tikz}
\definecolor{gold}{rgb}{0.85,.66,0}

\floatname{algorithm}{Algorithm}

\newcommand{\rbrac}[1]{\left( {#1} \right)}
\newcommand{\sbrac}[1]{\left[ {#1} \right]}
\newcommand{\cbrac}[1]{\left\lbrace {#1} \right\rbrace}
\newcommand{\round}[1]{\left\lfloor {#1} \right\rceil}

\newcommand{\norm}[1]{\left\Vert {#1} \right\Vert}
\newcommand{\eye}{\mathbf{I}}

\newcommand{\ones}{\mathbf{1}}

\newcommand{\expect}[1]{\mathbb{E}\left[ {#1} \right]}

\DeclareMathOperator*{\argmin}{argmin}

\newcommand{\abs}[1]{\left\vert {#1} \right\vert}

\newcommand{\mb}[1]{\mathbf{{#1}}}

\DeclareMathAlphabet\mathbfcal{OMS}{cmsy}{b}{n}

\newcommand{\xreal}{\mathbf{x}}

\newcommand{\Hreal}{\mathbf{H}}
\newcommand{\sreal}{\mathbf{s}}
\newcommand{\nreal}{\mathbf{n}}

\newcommand\BibTeX{{\rmfamily B\kern-.05em \textsc{i\kern-.025em b}\kern-.08em
T\kern-.1667em\lower.7ex\hbox{E}\kern-.125emX}}

\articletype{Original Research}%

\received{Jan 2018}
\revised{April 2018}
\accepted{May 2018}
\doi{10.1002/dac.3697}
\begin{document}
\title{Efficient Detection in Uniform Linear and Planar Arrays MIMO Systems under Spatial Correlated Channels}

\author[1]{João Lucas Negrão}
\author[2]{Taufik Abrão}

\authormark{João Lucas Negrão, Taufik Abrão*}

\address[1]{\orgdiv{DEEL -- Department of Electrical Engineering}, \orgname{UEL -- State University of Londrina}, \orgaddress{Rod. Celso Garcia Cid - PR445, Po.Box 10.011. CEP: 86057-970, Londrina, \state{Parana State}, \country{Brazil}}}

\corres{T. Abrão. Department of Electrical Engineering (DEEL), State University of Londrina (UEL). Rod. Celso Garcia Cid - PR445, 10.011. CEP: 86057-970, Londrina, Parana, Brazil. \email{taufik@uel.br}}

\abstract[Abstract]{In this paper, the efficiency of various MIMO detectors {was analyzed from} the perspective of highly correlated channels, where MIMO systems have a lack of performance, besides in some cases an increasing complexity. Considering this hard, but {a useful} scenario, various MIMO detection schemes {were accurately evaluated concerning} complexity and bit error rate (BER) performance. Specifically, successive interference cancellation (SIC), lattice reduction (LR) and the combination of them were associated with conventional linear MIMO detection techniques. {To demonstrate effectiveness}, a wide range of the number of antennas and modulation formats have been considered aiming to verify the potential of such MIMO detection techniques according to their performance-complexity trade-off. We have also studied the correlation effect when both transmit and receiver sides are equipped with uniform linear array (ULA) and uniform planar array (UPA) antenna configurations. The performance of different detectors is carefully compared when both antenna array configurations are deployed {considering} a different number of antennas and modulation order, especially {under} near-massive MIMO condition. We have also discussed the relationship between the array factor (AF) and the BER performance of both {antenna array} structures.}

\keywords{MIMO, Lattice Reduction, Channel Correlation, Zero-Forcing, MMSE, Sphere Decoding, ULA, UPA}


\maketitle


\section{Introduction}
The Multiple-Input Multiple-Output {systems are} recognized by the capacity to provide significant spectral efficiency and/or performance {enhancements} on wireless communication systems by the use of multiple antennas at both transmitter and receiver sides. In spatial multiplexing gain mode, the deployment of simultaneously transmit data streams through multiple antennas were developed to enhance the spectral efficiency at the cost of increasing data detection complexity at the receiver side \cite{wubbenLR}.
The V-Blast architecture, {proposed in the pioneer work \cite{VB}, {was} able to exploit the communication channel capacity,} providing spatial multiplexing gain and high data rates, which inspired so many works into multiple antenna systems. This spacial multiplexing gain on MIMO systems is achieved by dividing the total transmitted power over the antennas, taking advantage of the multi-path diversity to achieve a great array gain, in other words, more bits per second per Hertz of bandwidth are transmitted. Moreover, with {MIMO systems, improvements can be considered on the transmitted energy efficiency, data rates and/or symbol error rates, being defined by the antennas disposal at array configuration and the transmission-detection techniques applied.} In project meanings, it is necessary to balance these improvements with the available resources in the systems, this procedure is crucial, since energy and spectrum are a scarce resource. Thereby, the purpose is to provide solutions attaining to a performance improvement under a low or moderate complexity constraint. Hence, the goal of this work consists in study MIMO architectures equipped with low or moderate complexity detectors, keeping appropriate BER performance under full diversity condition. Moreover, linear MIMO detectors and their combinations with sub-optimal equalization techniques like ordering , interference cancellation (SIC) and LR were studied in terms of performance-complexity trade-off.

{Another relevant consideration in our work is the correlated fading channels; as currently the physical size of communication devices are being greatly reduced, the space to accommodate the antennas in those device is reducing as well. In realistic MIMO systems, operating under {ultra high frequency (UHF)} ranges, the desired antenna element spacing to provide an uncorrelated channel state is reasonably great. Moreover, MIMO systems equipped with a great number of antennas, and exploit the maximum multiplexing gain (or even the maximum diversity gain) is a project challenge.} {Thereby}, it is easy to conclude that a {correlated MIMO channel} scenario will cause degradation effects on the {performance, as well as the achievable rates;} and in practical conditions this will result in more transmitting power needed. {Hence, efficient MIMO detectors operating under proper BER performance and transmit power limits, which is directly connected to the SNR, are of great interest.}

Recently, large (or massive) MIMO systems have arrived as a {technology for 5G systems carrying many promises \cite{Boccardi}, such as higher spectral and energy efficiency and mainly the immunity to additive noise provided by very large arrays \cite{ScalingUp}.} When the number of antennas becomes large some effects arise, such as, channel properties that were random before now appears deterministic; e.g, singular values of the channel matrix approach to deterministic functions; system is limited by interference from other transmitters because thermal noise is averaged out \cite{ScalingUp}. Although, large arrays bring two main problems: correlation {between antennas and the signal processing complexity. The first one comes from the} fact that the accommodation area for the large arrays are small, causing the effect of the correlated channels. {The second occurs due to an increasing demand of signal processing which arises from the large number of antennas, which requires more hardware and operations from the system}. Therefore, the study on MIMO processing techniques is important to know the limitations of each scenario and to analyze the best choices, in terms of performance and complexity trade-off, to practical high efficiency communication systems. 

The decoupling of a transmitted signal originated from a received signal sample, can be designated as the main problem of MIMO detection. 
As it is known, the MIMO systems send data over different antennas, {that travel} over different paths, then the signal at the receiver side, at each antenna, is a combination of every transmit antenna signal and the received signal is a combination of every transmit antenna. There are many techniques on MIMO structures capable to decouple the transmitted signal, each one offers an achieved performance and a complexity level, the design challenge is to balance the available resources into the project requirements.

The optimal algorithm that achieves a minimum joint probability of error, detecting all the symbols simultaneously, is the maximum likelihood (ML) detector, that is known to be NP-hard. It can be carried out with a brute force-search over all possibilities in the transmitted vectors set, searching for the one that minimizes the Euclidean distance from the received vector. However, the expected computation complexity of the ML receiver is unpractical for many applications. Another possibility when considered looking for near-optimum performance is the sphere decoding (SD), that is {a promising approach on MIMO detection.} The SD provides lower complexity when compared to the ML for small noise value, but remains complex under low or medium SNR regions for real communication systems, becoming of the same order of ML complexity for low SNR region \cite{Jalden2,Barbero}.

Moreover, classic linear MIMO detection approaches are considered, such as the zero-forcing (ZF) detector which is know by being able to completely remove inter-antenna interference, at the cost of a significantly increase at the additive noise for ill conditioned channel matrices. There is also the minimum mean squared error (MMSE)-based detector which can be {considered} as a better alternative, once it considers the noise power {throughout the symbol detection procedure}.  Besides ZF and MMSE detectors when combined with SIC \cite{Bohnke} perform an layer-by-layer detection canceling the interference form the previous detected symbol. Since first layers detection errors can be propagated along the algorithm, the ordered SIC (OSIC) \cite{VB,WubbenQR} detector provides remarkable improvements on performance by detecting the most reliable antennas first. {Both ZF and MMSE detectors when combined with OSIC turns into detection schemes able to provide lower complexity compared to the ML or even the SD detector, however present greater degradation in the BER performance.} Furthermore, pre-processing techniques such as the lattice reduction (LR) \cite{RV,MaWei,Wubben} aided linear MIMO detectors can be used to simultaneously provide performance improvement and complexity reduction, since the transformed channel has quasi-orthogonality features it will improve the final quality of the detected signal, achieving, in some cases, near-ML performance. {The LR computational complexity is recognized as polynomial in time; however, highly correlated channel scenarios result in devastating impacts on the MIMO channel matrix estimation while the LR-aided linear MIMO detectors may result in an undesirable additional complexity, especially when the system is equipped with a high number of antennas \cite{RV}.} However, those problems are part of the challenge to implement the applicability of large-MIMO systems.

The contribution of this work are two fold: first, to provide a BER performance (reliability) {\it versus} complexity trade-off for a broad class of MIMO detectors operating under realistic scenarios, where we consider different antennas structures, specially ULA and UPA, under different correlated channels and system scenarios; second,  BER performance impact analysis when ULA and UPA correlated MIMO channels are deployed considering the array factor (AF) of each structure and its impact over the overall system performance.

The reminder of this paper is organized as follows. The spatial correlated MIMO channel modeling considering  linear and planar antenna arrays are is discussed in Section \ref{sec:sysmod}. Section \ref{sec:MIMO_Detec} revisits several well-known effective sub-optimum MIMO detectors, with the perspective to evaluate their performance under correlated channel with increasing number of antennas and different modulation orders. Section \ref{sec:perform} analyses the improvements and drawbacks of a collection of MIMO detectors operating under adverse scenarios in terms of channel correlation and antenna array configurations. Important complexity issues are addressed in Section \ref{sec:complexity}. Finally, Section \ref{sec:concl} offers the conclusion remarks.

\section{System Model}\label{sec:sysmod}
\begin{figure}[htbp!]
	\begin{center}	
	\includegraphics[width=.6\textwidth]{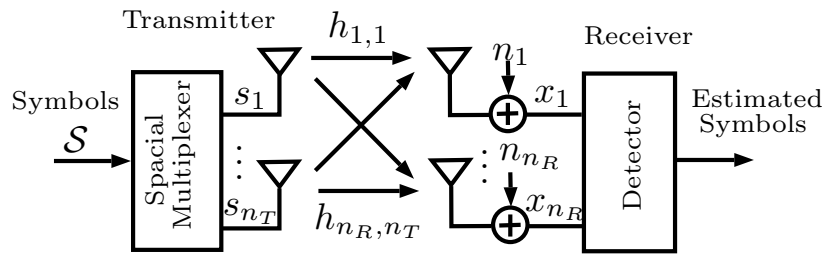}
	\caption{MIMO System with Spacial Multiplexing}
	\label{fig:MIMO_SYS}	
	\end{center}
\end{figure}

{Considering} a point-to-point MIMO system composed by $n_T$ transmit antennas and $n_R$ receive antennas, where the transmitted data is demultiplexed over the $n_T$ transmit antennas. A MIMO system topology is depicted in Fig.\ref{fig:MIMO_SYS}.

\vspace{-5mm}

The model is considered under an overdetermined MIMO system, i.e., $n_R \geq n_T$, working in spatial multiplexing mode. A classical problem in MIMO systems consists in reliably detect the transmitted symbol, despite the channel's distortion and noise \cite{larsson2009mimo}. Thereby, the received signal can be described by:
\begin{equation}\label{eq:MIMOsys}
	\xreal = \Hreal\sreal+\nreal, 
\end{equation}
where $\mb{s}_{n_T\times 1}$ symbols are transmitted through a channel which gain is represented by $\mb{H}_{n_R\times n_T}$ and additive noise $\mb{n}_{n_R\times 1}$. {Each element of matrix $\mb{H}$ represents the channel gain for a selected path and these gains are known at the receiver.} The {column-}vector {$\mb{x}_{n_R\times 1}$ represents the received  signal vector}, formed by the symbols after passing through the channel. 

It is also assumed that the noise vector $\mb{n}$, are samples of additive noise represented as circularly-symmetric Gaussian distribution, $\mb{n} \sim \mathcal{C}\mathcal{N}\{0,\sigma^2_n \mathbf{I}\}$, with variance $\sigma^2_n$. An alternative way to represent the noise statistics is through the covariance matrix $\expect{\mb{n}\mb{n}^H} = \sigma^2_n\eye_{n_R} $ 	

In order to achieve better spectral efficiency and performance we will consider a M-QAM modulation, where the symbols are denoted by a complex number limited to $\pm\rbrac{\sqrt{M}-1}$, in the real and imaginary part \cite{Bai,Wubben,Koba}.

The structure of the complex set is represented by 
{$$\mathcal{S} = \cbrac{a + jb\quad |\quad a,b \in \cbrac{-\sqrt{M}-1, -\sqrt{M}+3, \dots, \sqrt{M}-1}}$$}
For such modulation format, the average symbol energy is given by:
\begin{equation}
	E_s = \dfrac{2(M-1)}{3}
\end{equation}	

Also, it will be adopted Gray coded symbols, where adjacent symbols differ only one bit, which minimize the BER performance.

Furthermore, the channel model will be kept simple, however substantially adequate to the proposed systems. Specifically it is used a MIMO channel under Rayleigh fading and under the effect of spatial correlation between antennas. The Rayleigh fading is modeled as two random variables (r.v.) that follows circular complex Gaussian distribution, with zero-mean and unitary variance, i.e., $h_{ij} \sim \mathcal{C}\mathcal{N}\{0,1\}$, whose magnitude is represented by a Rayleigh r.v., while the phase is represented by a uniform distributed r.v. \cite{Cho:2010}. Furthermore, it is worth noting that the Rayleigh fading model is valid for environments that are rich in scattering, i.e., highly urbanized environments or with great number of obstacles. {It} means that the signal do not have a prevalence path, i.e., non-line-of-sight (NLOS) channels \cite{Goldsmith:2005}.

\subsection{Correlated MIMO Rayleigh-Fading Channels} \label{Corr}
This section discusses the MIMO channel correlation among different antenna structures. As we have already defined the channel basic characteristics, the next step is to evaluate the correlation effect and how to model it. The space for the accommodation of antennas elements in wireless systems is in many cases limited. Thus, the correlation of antennas appears as an aggravating fact in MIMO systems, and especially in large-scale MIMO systems. As the correlation between antennas increase, the channel between them {gets} more similar to each other and this is caused by decreasing the distance between antennas. Generally, {the channels start to present correlation when the distance between antennas is lower then a half wave length $\rbrac{\lambda/2}$ \cite{Goldsmith:2005}. With highly correlated channels, {spatial} diversity {loss} is expected and consequently, deterioration in system performance and capacity.}  

\begin{figure}[!htbp]
\centering
\includegraphics[width=0.26\textwidth]{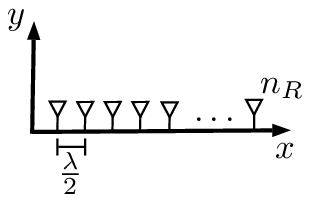}\,\,\,\,\qquad 
\includegraphics[width=0.28\textwidth]{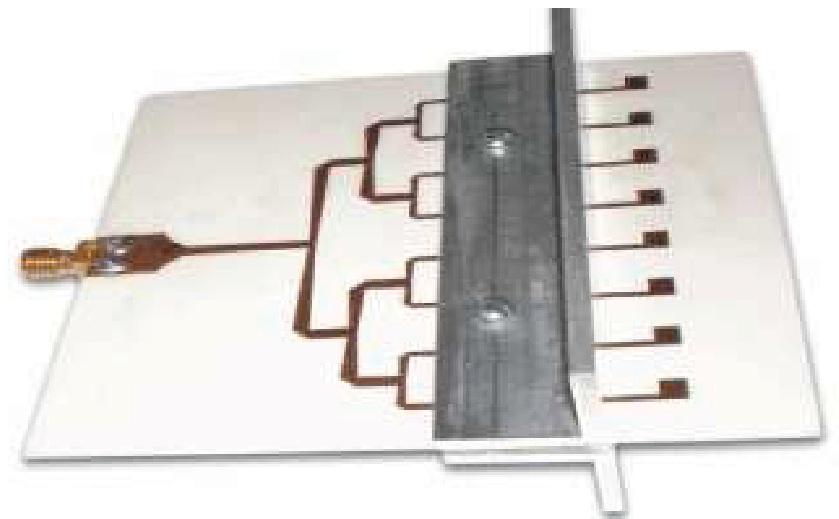}\\
\vspace{-1mm}
a) \hspace{50mm} b)
\vspace{-1mm}
\caption{a) Uniform Linear Array (ULA) model; b) Example of an ULA implementation; Photo Source: \cite{ULA_real}}
\label{fig:ULA}
\end{figure}

Commonly, the classical and simple configuration {allowing} us to analyze correlation for MIMO systems is the one where the distribution is organized as an uniform linear array(ULA) \cite{van}, Fig. \ref{fig:ULA}, which simplifies the antenna model while allows a very good approximation for the correlation effect at MIMO systems with a low or moderate number of antennas. On the other hand, when the number of antennas are considerably increased, {\it i.e.} massive MIMO applications, another array structures are required in order to accommodate the transmit antenna elements in a feasible way. Different array possibilities and configurations have been proposed for the large MIMO channel; as a consequence, different correlated Massive MIMO channel models have arisen.

\begin{figure}[!htbp]	
\centering
\includegraphics[width=0.23\textwidth]{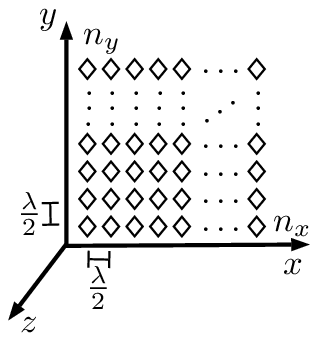}\qquad\qquad\qquad
\includegraphics[trim={1.3cm 1cm 2cm 2cm}, clip,width=0.15\textwidth]{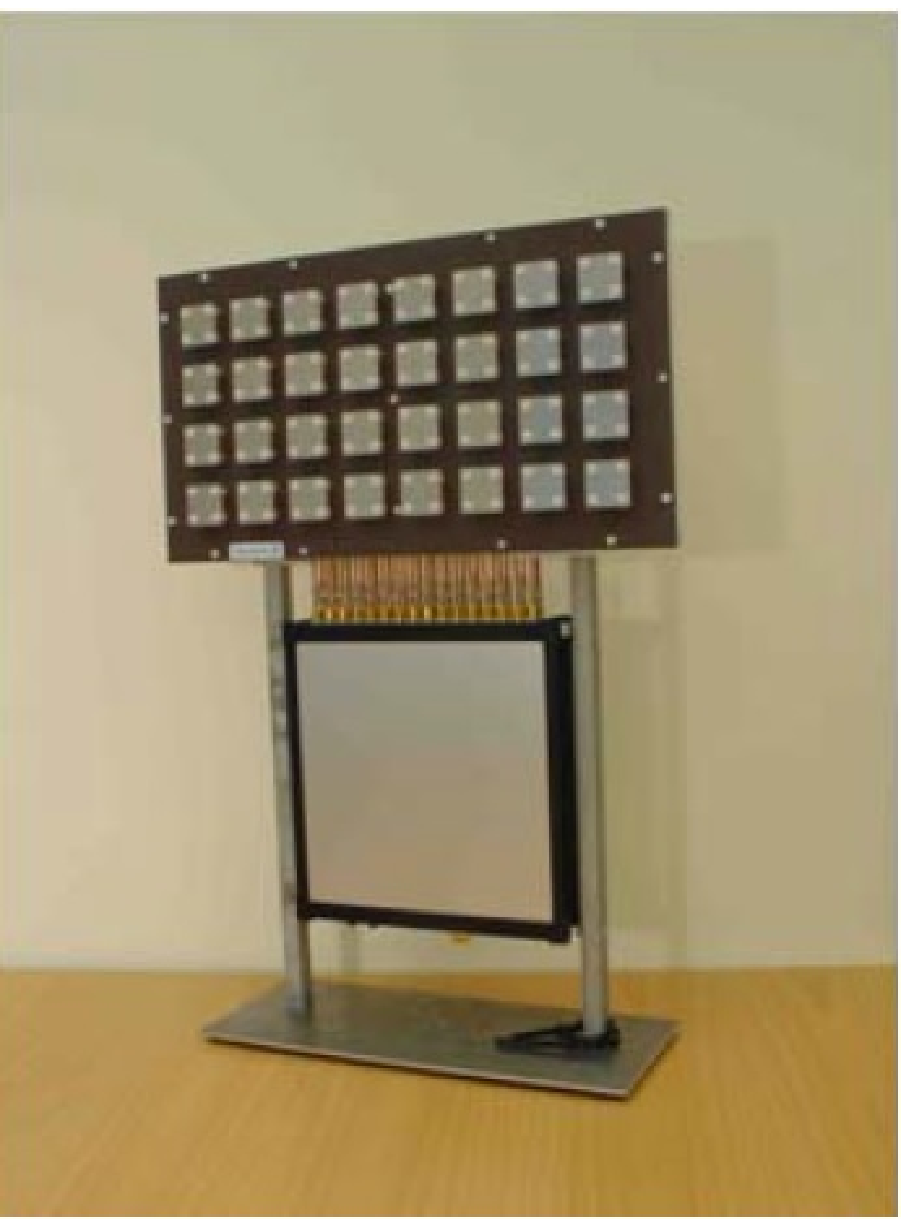}\\
\vspace{-1mm}
a) \hspace{60mm} b)
\vspace{-1mm}
\caption{a) Uniform Planar Array (UPA); b) Example of an UPA implementation. Photo Source: \cite{UPA_real}}
\label{fig:UPA}
\end{figure}

One of the first proposed antenna array {arrangement is} the uniform planar array (UPA), In Fig. \ref{fig:UPA}, antenna elements are disposed in a two dimensional array. {Accordingly to \cite{Balanis}, planar array structures supply additional variables which can be used to control and shape the pattern array. Also, providing more versatility allowing more symmetrical patterns with lower side lobes at the total radiated power pattern. Additionally, they can be used as a scan mechanism for the main beam of the antenna towards any point in space.} 

In our work it will be considered two antenna array structures. The classic and simple ULA configuration will be taken as reference, and the UPA array, which can be identified as a promising candidate to compose the {base station (BS)} antenna structure in massive MIMO scenarios, will be considered as well. Those choices were made aiming to evaluate the UPA implementation impact at the BS, because theoretically the planar array structure has the potential to concentrate the downlink beamforming at the transmitter side. This characteristic can be showed through the Array Factor (AF), which is the factor by which the directivity function of an individual antenna must be multiplied to get the directivity of the entire array.

According to the antenna theory, the array factor of an ULA of $N$ elements along the $x$-axis can be represented as \cite{Balanis}:
\begin{equation} \label{eq:AF_ula}
	AF_{\rm ula} = \sum_{n = 1}^{N} I_n e^{j(n-1)(kd\sin\theta\cos\phi)}
\end{equation}
where, $\sin\theta\cos\phi$ is the directional cosine with respect to the $x$-axes, $I_n$ is the amplitude excitation factor of each element and $d$ is antenna element spacing. For simplification purposes, equation \eqref{eq:AF_ula} can be written as:
\begin{equation} \label{eq:AF_ula_simp}
	AF_{\rm ula} = \sum_{n = 1}^{N} I_n e^{j(n-1)\psi}
\end{equation}
where $\psi = (kd\sin\theta\cos\phi)$ {and $k = \frac{2\pi}{\lambda}$.}

According to \cite{Balanis}, the AF in \eqref{eq:AF_ula_simp} can be expressed in an alternate, compact and closed form whose function and their distribution are more recognized. This is accomplished by multiplying both sides of \eqref{eq:AF_ula_simp} by $e^{j\psi}$, then we have:
\begin{equation} \label{eq:AF_ula_open}
	(AF_{\rm ula})e^{j\psi} = e^{j\psi} + e^{j2\psi} + e^{j3\psi} + \dots + e^{j\rbrac{N-1}\psi} + e^{jN\psi}
\end{equation}
Subtracting \eqref{eq:AF_ula_simp}  from \eqref{eq:AF_ula_open} reduces to
\begin{equation}
	(AF_{\rm ula})\rbrac{e^{j\psi} - 1} = \rbrac{-1 + e^{jN\psi}}
\end{equation}
which can also be written as
\begin{equation}\label{eq:AF_ula_ff}
	AF_{\rm ula} = {\sbrac{\dfrac{e^{jN\psi }- 1}{e^{j\psi} - 1}} }  = e^{j[(N-1)/2]\psi}\sbrac{\dfrac{\sin\rbrac{\frac{N}{2}\psi}}{\sin\rbrac{\frac{1}{2}\psi}}}
\end{equation}
and according to \cite{Balanis}, if we take as reference point the physical center of the array, the AF of \eqref{eq:AF_ula_ff} reduces to
\begin{equation} \label{AF_fff}
	AF_{\rm ula} = \sbrac{\dfrac{\sin\rbrac{\frac{N}{2}\psi}}{\sin\rbrac{\frac{1}{2}\psi}}}
\end{equation}
In general lines, the array factor can be represented as a function of the number of elements, their geometrical disposal, corresponding magnitude, relative phases and element spacing. With those considerations, the AF should result in a simpler form if each element have identical amplitude, phase, and spacing related each other, which motivates a normalization in the AF expression, providing a fair comparison for different arrangements.

Substituting $\psi$ into \eqref{AF_fff} and considering $I_n = 1$  we have the normalized version of AF for ULA, which is expressed as:
\begin{equation}
	AF_{\rm ula}(\theta,\phi) = \dfrac{1}{N}\dfrac{\sin(N\frac{kd\sin\theta\cos\phi}{2})}{\frac{kd\sin\theta\cos\phi}{2}}
\end{equation}
which represents the directivity pattern of the ULA.

Now if $L = \frac{N}{2}$ antenna elements are placed in the $x$-axes and in the $y$-axes, a rectangular/planar array will be formed. Assuming again that all elements are equally spaced with intervals $d_x$ and $d_y$ in both axes, and all elements have the same amplitude excitation $I_{l}$, the UPA array factor can be represented as:
\begin{equation}
	AF_{\rm upa} = I_{l} \sum_{l = 1}^{L} e^{j(l-1)(kd_x\sin\theta\cos\phi)}\sum_{l = 1}^{L} e^{j(l-1)(kd_y\sin\theta\sin\phi)}
\end{equation}
the normalized UPA array factor can be obtained as:
\begin{equation}
	AF_{upa}(\theta,\phi) = \cbrac{\dfrac{1}{L}\dfrac{\sin(L\frac{kd_x\sin\theta\cos\phi}{2})}{\frac{kd_x\sin\theta\cos\phi}{2}}}\cbrac{\dfrac{1}{L}\dfrac{\sin(L\frac{kd_y\sin\theta\sin\phi}{2})}{\frac{kd_y\sin\theta\sin\phi}{2}}},
\end{equation}
where $k = \frac{2\pi}{\lambda}$.

\begin{figure}[!htbp]
\centering 
\includegraphics[width=0.99\textwidth]{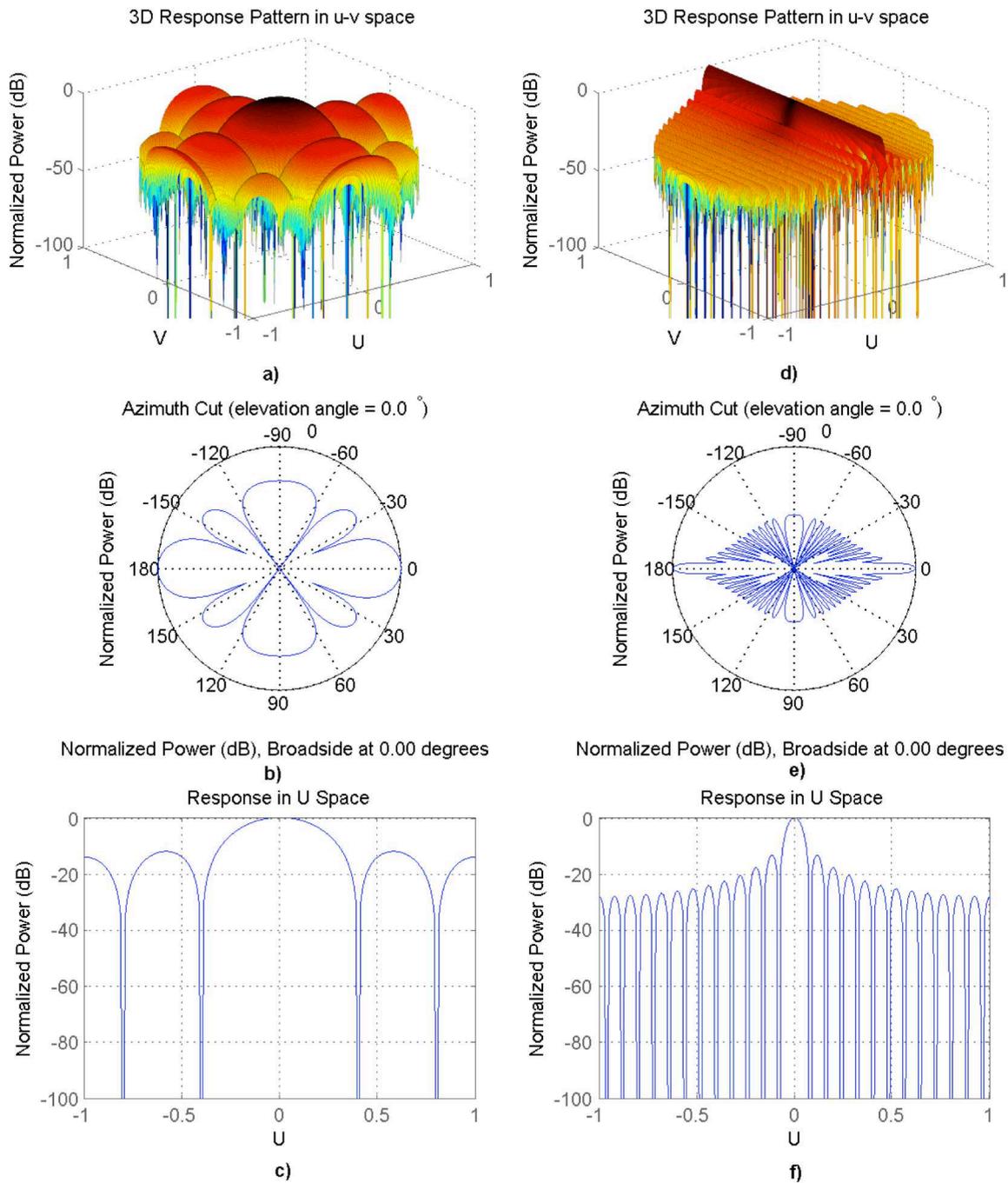}
\vspace{-12mm}
\caption{Array Factor for $0.5 \; \lambda$ element-spaced: [left] UPA of $5 \times 5$ elements;  \qquad [right]: ULA with $25$ elements}
\label{fig:UPA_ULA_AF}
\end{figure}

The gain inside the array factor of a $5 \times 5$ UPA and $25$ elements ULA {has} been plotted in Figure \ref{fig:UPA_ULA_AF} aiming to identify their beam pattern and normalized power distribution over the azimuth and elevation directions, which directly impact on the array gain.

Further elaboration is depicted in Fig. \ref{fig:UPA_ULA_AF}, which is the case where the element spacing is defined as $d=0.5 \; \lambda$, {and a frequency of $1$GHz}. The normalized beam pattern in polar coordinates and a cross section in the U-plane, where the normalized energy distribution is plotted as a function of the elevation angle variations projected onto the Cartesian plane, is also described. This coordinates projection over the Cartesian plane is known as UV mapping and it is commonly used in antenna theory, image processing and also 3D drawing. The UV mapping is a $\mathbb{R}^3$ to $\mathbb{R}^2$ projection which transform a 3D pattern on its 2D rectangular projection.

The $u-v$ coordinates can be easily derived from the $\phi$ and $\theta$ angles which are respectively the azimuth and elevation angles in spherical coordinates. 
The relationship between these two coordinates system is simply:
\begin{equation} \label{eq:project}
\arraycolsep = 1.4pt
\begin{array}{rcl}
u &=& \sin\theta\cos\phi \\
v &=& \sin\theta\sin\phi
\end{array}
\end{equation}
the values of $u$ and $v$ satisfy the inequalities:
\begin{equation}
\arraycolsep = 1.4pt
\begin{array}{rcl}
-1 \leq u &\leq& 1 \\
-1 \leq v &\leq& 1 \\
u^2 + v^2 &\leq& 1
\end{array}
\end{equation}

Fig. \ref{fig:UPA_ULA_AF} a) and d) represent the 3D pattern in terms of normalized power for both UPA and ULA, respectively. It is simple to notice that the UPA structure presents a wider beam-width at the main lobe which provides larger beam gains that leads, at the BS, to lower transmit power and larger antenna coverage. On the other hand, the ULA power distribution is more heterogeneous presenting smaller beam-width and, a power concentration directly at the beam direction which provide a smaller coverage area with the total power transmitted. Another observed characteristic due to the UPA structure deployment, is that it provides larger side lobes when compared to the ULA side lobes, which implies in more transmit gain at the UPA side lobes benefiting the power distribution with this array structure. In order to manipulate the antenna beam pattern, especially the beam-width; there are {two variables} in the array structure that can modify the pattern distribution. The first is the number of antenna elements, which directly affects the beam-width, so that with increasing number of elements, the main lobe beam-width tends to become concentrated and the side lobes will suffer from higher attenuation. Another parameter impacting the beam pattern is the antenna element spacing; by decreasing the space between elements the beam-width become wider, {providing higher normalized power along the array, directly impacting in less attenuation at the transmitted signal.}

Thereby, the performance of MIMO systems equipped with both UPA and ULA arrays can be analyzed in a comparative meanings, aiming to determinate which is the best scenario to apply the planar array structure in comparison to the classical ULA approach. {The following sections {provide} mathematical expressions which {represent} the spacial correlation function for both studied antenna array structures, also providing a comparison between the simplified version and the full geometrical correlation matrix for each structure}.

\subsection{Uniform Linear Array} \label{sec:ULA}
{Several} MIMO channel correlation models {were} proposed in the last decades; one of the most important yet simple class of MIMO channel models is the one that assume independence among the correlation between transmit antennas (TX) and receive antennas (RX) (and vice versa). Hence, a spatially correlated MIMO fading channel is decently modeled by flat Rayleigh distribution and the correlation among antennas elements will be determined over the Kronecker's correlation model \cite{van}, as follows:
\begin{equation}\label{chap2:corrULA}
\mb{H} = \sqrt{\mb{R}_{H,Rx}}\mb{H'}\sqrt{\mb{R}_{H,Tx}}
\end{equation}
where $\mb{H'} (n_R \times n_T)$ is the uncorrelated MIMO channel which is represented with independent, identically distributed (i.i.d.) complex Gaussian with zero-mean and {unitary variance}, $g_{ij} \sim \mathcal{C}\mathcal{N}\{0,1\}$. The correlation matrices $\mb{R}_{H,Tx} (n_T \times n_T)$ and $\mb{R}_{H,Rx}(n_R \times n_R)$ denote the spatial channel correlation held among the transmitter and receiver side, respectively. Each element of those matrices are represented, in terms of a normalized correlation index $\rho$, by:
\begin{equation}
\arraycolsep = 1.4pt
\left\{\begin{array}{rcl}
r_{H,Rx \; ij} & = & \rho^{(i-j)^2} \\
r_{H,Tx \; ij} & = & \rho^{(i-j)^2}
\end{array}\right.
\end{equation}

Note that matrix $\mb{H'}$ in \eqref{chap2:corrULA} is similar to matrix $\mb{H}$. Hence, for the rest of this work we assume that the Tx and Rx antenna elements are {equidistant}, with identical number of antennas {$n_T = n_R = n$} and {consequently the same} correlation matrix $\mb{R}_{H,Rx} = \mb{R}_{H,Tx} = \mb{R}_{H}$, {that is represented as}:
\begin{equation}
\mb{R}_H = \begin{bmatrix}
1 & \rho & \rho^4 & \dots & \rho^{(n-1)^2}\\
\rho & 1 & \rho & \dots & \vdots \\
\rho^4 & \rho & 1 & \dots & \rho^4 \\
\vdots & \vdots & \vdots & \ddots & \rho \\
\rho^{(n-1)^2} & \dots & \rho^4 & \rho & 1 \\
\end{bmatrix}
\end{equation}%
Also note that a fully uncorrelated channel means $\rho = 0$, while an entirely correlated scenario results in $\rho = 1$.

\subsection{Uniform Planar Array} \label{sec:UPA}
Traditionally, the MIMO systems adopt ULA setup as the simplest and standard structure. But considering the used space limitation, the ULA setup is not suitable for large-scale antenna arrays. Hence, for massive MIMO applications the necessity to adopt a two-dimensional array structure, such as UPA, is essential. A correlation matrix for the UPA structure was proposed by \cite{URA_model}. In this paper a multidimensional array correlation structure is constructed for the UPA, based on a Kronecker product of two ULA correlation matrices. More specifically, considering a UPA constructed with isotropic antenna elements lying on the {\it XY} plane with $n_x$ and $n_y$ antenna elements along $x$ and $y$ coordinates, respectively, so that $n_r = n_x\cdot n_y$.

Moreover, we can assume an approximation in which the correlation between elements along $x$ coordinate does not depend on $y$ and is given by matrix $\mb{R}_{H,x}$, and the correlation along $y$ coordinate does not depend on $x$ and is given by matrix $\mb{R}_{H,y}$. The following Kronecker-type approximation of the UPA correlation matrix is proposed by \cite{URA_model}:

\begin{equation}\label{eq:UPA}
\arraycolsep = 1.4pt
\begin{array}{rcl}
\mb{R}_{H,r}  &=& \mb{R}_{H,x}\otimes \mb{R}_{H,y}\,\,
=\,\,\,\,\begin{bmatrix}
r_{H,x \; 1,1}\mb{R}_{H,y}& \dots & r_{H,x \; 1,n_T}\mb{R}_{H,y} \\ 
\vdots& \ddots & \vdots\\ 
r_{H,x \; n_R,1}\mb{R}_{H,y}& \dots & r_{H,x \; n_R,n_T}\mb{R}_{H,y} 
\end{bmatrix}
\end{array}
\end{equation}
where $\otimes$ denotes the Kronecker product. The equation \eqref{eq:UPA} indicates that the UPA correlation matrix $\mb{R}_{H,r}$ is the Kronecker product of two ULA correlation matrices $\mb{R}_{H,x}$ and $\mb{R}_{H,y}$, which are Toeplitz. Therefore, according to the authors, even tough $\mb{R}_{H,r}$ may not be a Toeplitz matrix, its approximation \eqref{eq:UPA} has a Toeplitz per block structure. According to \cite{URA_2}, this approximation model is reasonably accurate, allowing the usage of the well-developed theory of Toeplitz matrices for the analysis of multidimensional antenna arrays.

\subsection{Geometrical Correlation Model}
\begin{figure}[!htbp]
	\begin{center}
		\includegraphics[width=0.25\textwidth]{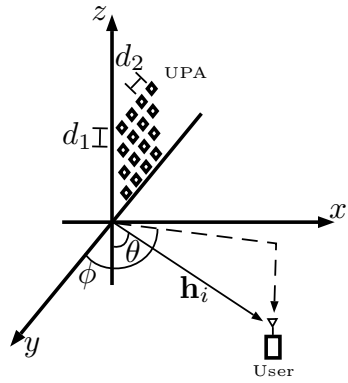}
		\vspace{-2mm}
		\caption{3D Channel model adopted to derive the spacial correlation function. $\phi$ is the azimuth angle, and $\theta$ is the elevation angle.}
		\label{fig:UPAcorr}
	\end{center}
\end{figure}

Another perspective to derive the correlation expression is made through the geometric properties of the problem. In \cite{Ying} the UPA analytical expression was derived based on a 3D channel model. The spacial correlation function was derived in a downlink transmission, where the BS is equipped with $N_v$ vertical antenna elements spaced by $d_1$ wavelengths, and $N_h$ horizontal antennas with $d_2$ wavelength spacing separation, as sketched in Figure \ref{fig:UPAcorr}.

The $(a,b)$-th antenna element denotes the antenna in the $a$-th row and the $b$-th column of the UPA, so the channel from the $(a,b)$-th element in the transmitter to the receiver is associated with the $b + N_h(a-1)$-th element of $\mb{h}_i$, which is the channel vector related to the $i$-th fading block. As modeled by \cite{ZhaoWang}, the spacial correlation matrix $\mb{R}_h$ for the UPA is composed by the correlation element between the $(a,b)$-th and $(p,q)$-th antennas, given as:
\begin{equation}\label{eq:UPACorr}
[\mb{R_h}]_{(a,b),(p,q)} = \frac{D_1}{\sqrt{D_5}}e^{-\frac{D_7 + (D_2(\sin\phi)\sigma)^2}{2D_5}}e^{j\frac{D_2 D_6}{D_5}}
\end{equation}
with the variables defined by:
\begin{equation}\label{eq:UPACorr_Var}
\arraycolsep = 1.3pt
\begin{array}{rcl}
D_1 &=& e^{j\frac{2\pi d_1}{\lambda}(p-a)\cos\theta}e^{-\frac{1}{2}(\xi\frac{2\pi d_1}{\lambda})^2(p-a)^2\sin^2\theta}, \\[0.5em]
D_2 &=& \dfrac{2\pi d_2}{\lambda}(q-b)\sin\theta, \\[0.5em]
D_3 &=& \xi\dfrac{2\pi d_2}{\lambda}(q-b)\cos\theta, \\[0.5em]
D_4 &=& \dfrac{1}{2}(\xi\dfrac{2\pi}{\lambda})^2 d_1d_2(p-a)(q-b)\sin(2\theta), \\[0.5em]
D_5 &=& (D_3)^2((\sin\phi)\sigma)^2+1, \\[0.5em]
D_6 &=& D_4((sin\phi)\sigma)^2 + \cos\phi, \\[0.5em]
D_7 &=& (D_3)^2\cos^2\phi - (D_4)^2((\sin\phi)\sigma)^2 - 2D_4\cos\phi,\\
\end{array}
\end{equation}
where $\lambda$ is the carrier wavelength, $\phi$ is the azimuth angle-of-departure (AoD), $\theta$ is the elevation AoD, while $\sigma$ is the standard deviation of horizontal AoD, and $\xi$ is the standard deviation of vertical AoD. Finally, $\mb{R}_h$ is a $n_T \times n_T$ matrix and $[\mb{R}_h]_{(a,b),(p,q)}$ is the element at the $b + N_v(a-1)$-th row and the $q + N_h(p-1)$-th column.

\vspace{-5mm}
\vspace{-7mm}

Considering the above analytical expressions, it is clear that term $D_1$ is only associated with the elevation angle, containing only $(p-a)$ terms, while $D_2,D_3$ and $D_5$ are azimuth related containing only the $(q-b)$ terms. Variables $D_4,D_6$ and $D_7$ have the cross term $(p-a)(q-b)$, containing both elevation and azimuth correlations. However, $D_6$ and $D_7$ are functions of $D_4$. As proposed by \cite{Ying}, if term $D_4$ could be neglected, i.e, $D_4 = 0$, the correlation term $[\mb{R}_h]_{(a,b),(p,q)}$ can be written as a simple product of elevation and azimuth correlations. Therefore, if $D_4 = 0$, the correlation matrix is separable:
\begin{equation} \label{eq:geoUPA}
\mb{R}_{h} = \mb{R}_{az}\otimes\mb{R}_{el},
\end{equation}
where the elements of elevation correlation matrix are expressed as:
\begin{equation}
[\mb{R}_{el}]_{(a,b)} = e^{j\frac{2\pi d_1}{\lambda}(p-a)\cos\theta}e^{-\frac{1}{2}(\xi\frac{2\pi d_1}{\lambda})^2(p-a)^2\sin^2\theta}
\end{equation}
and the correlation elements in the azimuth direction are:
\begin{equation}
[\mb{R}_{az}]_{(p,q)} =  \frac{1}{\sqrt{D_5}}e^{-\frac{D^2_3\cos^2\phi}{2D_5}}e^{j\frac{D_2\cos\phi}{D_5}}e^{-\frac{(D_2(\sin\phi)\sigma)^2}{D_5}}
\end{equation}

It is demonstrated by \cite{Ying} that the Kronecker correlation model has very similar eigenvalues distribution as the correlation matrix, and thus is a good approximation for the original UPA correlation matrix. 

The equation that express the ULA spacial correlation function is derived at \cite{Buehrer}, and is defined as:
\begin{equation}\label{eq:geoULA}
[\mb{R}_{\rm ula}]_{(i,j)} = e^{j\frac{2\pi d}{\lambda}(i-j)\sin\theta}e^{-\frac{1}{2}(\xi\frac{2\pi d}{\lambda})^2(i-j)^2\cos^2\theta}
\end{equation}
where $d$ is the distance between antenna elements. This expression is similar to the elevation correlation, $[\mb{R}_{\rm el}]_{(a,b)}$, in the UPA structure. From the previous expressions it is easy to conclude that the ULA and UPA models provided in subsection \ref{sec:ULA} and \ref{sec:UPA}, respectively, are the simplified expressions to the above geometrical models. 

As we have derived the correlated channel expressions for both antenna array structures, now we will introduce several MIMO detectors that will be deployed to detect the transmitted symbols from the BS to the {mobile terminal (MT)} (downlink); hence, the BS antenna correlation effect plays an important role on the system capacity/reliability reduction.

\section{{MIMO Detection Techniques}}\label{sec:MIMO_Detec}
The present section recall the commonly approaches for MIMO detectors techniques, going through the maximum-likelihood (ML), sphere decoder (SD), zero-forcing (ZF) and minimum mean squared error (MMSE). Also, it will be provided a succinct discussion over the application of {two  techniques applicable to the MIMO detection context, {\it i.e.,}  the sucessive} interference cancellation (SIC) and lattice reduction (LR).

\subsection{Maximum Likelihood (ML)}
The ML detector perform an exhaustive search over the whole set of possibles symbols $s \in \mathcal{S}^{n_T} $, of size $M^{n_T}$, in order to decide in favor of the one that minimizes the Euclidean distance, and therefore the lowest error, between the received signal $\mb{y}$ and the reconstructed signal $\mb{Hs}$:
\begin{equation}
	\widehat{\mb{s}} = \displaystyle \argmin_{s \in \mathcal{S}^{n_T}} \norm{{\mb{x}} - \mb{Hs}}^2.
\end{equation}
It is well known that the ML detector ensures the lowest BER performance in all the spectrum of MIMO detectors, but the search complexity grows exponentially according to the number of antennas and the number of symbols. {If we consider} a $M$-ary modulation with $n_T$ transmit antennas, each one transmitting a {different} symbol {in a distinct} time-slot system, the {order} of combinations is given by $M^{n_T}$.  It becomes impractical in cases where the constellation order and the number of antennas are considerably increased; for $M = 16$ and $n_T = 8$, the number of candidates becomes incredibly large, over $\approx 4$ billion of candidate-symbols must be evaluated.
%

\subsection{Sphere Decoder (SD)} \label{sec:SD}
Pursuing a reduction in the ML complexity, a similar approach has been proposed, namely the sphere-decoder (SD) detector, {that searches only the candidates bounded in the hypersphere of radius $d$, causing it performance to be highly related to the SNR:}
\begin{equation}
	d^2 < \norm{{\mb{x}} - \mb{H}\mb{s}}^2
\end{equation}
{If the search radius is too high, the SD complexity get close to the ML one. In contrast, if the search radius is set too small, no candidate will be chosen upon hypersphere.} Moreover, in order to obtain candidate-solution points to perform the sphere detection, it is necessary rewrite the eq.\eqref{eq:MIMOsys}, evaluating the QR decomposition of the channel matrix such as $\mb{H} = \mb{Q}\mb{R}$. The is evaluation will result in an orthogonal matrix $\mb{Q}$, with $\eye = \mb{Q}^H\mb{Q}$, and an upper triangular matrix $\mb{R}$; considering the detection purposes {those matrix have very} convenient properties. The procedure is performed as follows:
\begin{equation}
	\arraycolsep = 1.4pt
	\begin{array}{rcl}
		\mb{y} & = & {\mb{Q}^{H}} \mb{x}\,\,
		 =  \,\, \mb{Q}^H\mb{Q}\mb{R}\mb{s} + \mb{Q}^{H} \mb{n}\,\,
		 = \,\, \mb{R}\mb{s} + \mb{n}'.
	\end{array}
\end{equation}

Since the matrix $\mb{Q}$ is orthogonal, the statistical properties of the additive noise, $\mb{n}'$, remains unaltered and no noise {increment is foreseen}. Moreover, {as} $\mb{R}$ is an upper triangular matrix it enables noise estimation for each antenna independently. Hence, the points inside the hyper-sphere can be determined layer-by-layer, starting from the last row of $\mb{R}$, by evaluating:
\begin{equation} \label{eq:SD}
	d^2 < \norm{\mb{y} - \mb{R}\mb{s}}^2
\end{equation}

Considering {$\mb{R} = \begin{bmatrix} \mb{r}_1^T & \mb{r}_2^T & \mb{r}_3^T & \dots & \mb{r}_{n_T}^T 
\end{bmatrix}$}, the noise norm is given by:
\begin{equation} \label{Noise_norm}
	\norm{\mb{n}'}^2 = \norm{\mb{y} - \mb{R}\mb{s}}^2 = \sum_{k = 1}^{n_T}\abs{y_k - \mb{r}_k\mb{s}}^2.
\end{equation}

In fact, eq. \eqref{Noise_norm} shows that the noise norm is the summation of each layers noise independently. This way, the noise norm can be updated as the symbols are tested in each layer, which avoids the evaluation of the estimated noise for every symbol combination.

A beneficial feature that emerges from the structure of the detection problem in \eqref{eq:SD} is that, due to the upper triangular properties of the $\mb{R}$ matrix, {the tree search algorithm scan the symbol vector backwards, starting from the last antenna symbol to the first one, testing all candidates symbols recursively and independently, contrarily the ML.} {As this layer-by-layer procedure follows the radius restriction defined in \eqref{Noise_norm}, by finishing the SD detection, the most likely symbol-vector bounded by the hypersphere of radius $d$ is the solution.}

\subsection{Zero-Forcing (ZF)} \label{sec:ZF}
The Zero-Forcing detector, is a simple linear MIMO receiver, with low computational complexity. It is designed to suppress channel interference by multiplying the signal received by the Moore-Penrose pseudo-inverse of the channel matrix:
\begin{equation}
	\mb{H}^{\dagger} = \rbrac{\mb{H}^{H}\mb{H}}^{-1}\mb{H}^H.
\end{equation} 
With that, the estimated signal from the detector can be determined by:
\begin{equation}
	\widehat{\mb{s}} = \mb{H}^{\dagger}{\mb{x}} = \mb{s} + \mb{H}^{\dagger} \mb{n} \label{ZF}
\end{equation}
Considering a scenario without noise, the ZF detector has an identical ML performance, due to all channel interference suppression. Otherwise, in noise scenarios, ZF leads to noise enhancement. That problem inhibits the performance of the ZF algorithm due to ill-conditioned $\mb{H}$ matrices i.e near to {linearly dependent columns condition}, which after the {matrix inversion} in \eqref{ZF} leads to enhancements in the thermal noise variance in $\widehat{s}$ when compared to $\mb{y}$ \cite{Mirsad}.

\subsection{Minimum Mean Squared Error (MMSE)} \label{sec:MMSE}
The MMSE detector can be seen as a particularly useful extension of the ZF detection, which by taking the noise and signal statistics into account the detector is able to improve the overall MIMO detection performance. The procedural difference to MMSE is that, instead of the pseudo-inverse, MMSE uses:
\begin{equation}
	\mb{H}^{\dagger} = \rbrac{\mb{H}^{H}\mb{H} + \sigma^2_n \eye_{n_T}}^{-1}\mb{H}^H.
\end{equation} 
And the solution of the MMSE detector is:
\begin{equation}
	\widehat{\mb{s}} = \rbrac{\mb{H}^{H}\mb{H} + \sigma^2_n \eye_{n_T}}^{-1}\mb{H}^H {\mb{x}}. \label{MMSE}
\end{equation}
In another perspective, the MMSE detection can be {fulfilled} as:
\begin{equation}
	\arraycolsep = 1.4pt
	\begin{array}{rcl}
		\widehat{\mb{s}} & = & \underline{\mb{H}}^\dagger \underline{{\mb{x}}}\,\,
		 = \,\, \mb{s} + \underline{\mb{H}}^{\dagger} \mb{n}\\
	\end{array}
\end{equation}
It is easy to note that the equation above has the same structure of \eqref{ZF}, but the vector signal and the received vector are extended and respectively given by:
\begin{equation}
	\underline{\mb{H}} = \begin{bmatrix}
		\mb{H}  \\
		\sigma_n \eye_{n_T}	
	\end{bmatrix}, \qquad \qquad {\underline{\mb{x}}} = \begin{bmatrix}
		{\mb{x}} \\
		\mb{0}_{{n_T} \times 1}
	\end{bmatrix}
\end{equation} 
The extended matrix model is more complex than the approach given in \eqref{MMSE}, but this model is required on successive interference cancellation (SIC) and can be used on lattice-reduction in order to achieve performance improvements \cite{Wubben}.

\subsection{Successive Interference Cancellation (SIC)} \label{sec:SIC}
The SIC detection technique can be {performed} by evaluating the QR decomposition of the matrix $\mb{H}$, which was addressed in section \ref{sec:SD}. An important observation is due to the fact that for ZF {detectors}, the QR decomposition should be {executed} on $\mb{H}$, while for MMSE {cases}, it is applied at the $\underline{\mb{H}}$ matrix. As we already know, the MIMO detection aided QR can be performed as {follows}:
\begin{equation}
	\widehat{\mb{s}} = 	\mb{Q}^H {\mb{x}} = \mb{Rs} + \mb{Q}^H \mb{n}.
\end{equation}

Since $\mb{Q}$ is an orthogonal matrix, when multiplied by the noise term, $\mb{Q}^H \mb{n}$, the statistical properties of the additive noise remains unaltered. As matrix $\mb{R}$ has an upper triangular structure, the $n$-th element of $\widehat{\mb{s}}$ is completely free of inter-antenna interference, and can be used to correctly estimate the received signal after the addition of an appropriate scale of $\dfrac{1}{r_{ii}}$, where, $i = n_T$ \cite{WubbenQR}. Hence, the linear system can be solved upwards by:
\begin{equation}
	\arraycolsep = 1.4pt
	\widehat{\mb{s}}  = \left\{	\begin{array}{ll}
		{\dfrac{x_i}{r_{ii}},} &  i = n_T\\
		\dfrac{1}{r_{ii}}\rbrac{{x_i} - \displaystyle \sum_{k =i+1}^{n_T}r_{ik}\widehat{s_k}}, &  i = n_T - 1, \dots, 3,2,1 \\
	\end{array}\right. \label{SIC}
\end{equation}
{It is important to note that each symbol must pass to the slicer before following to the interference cancellation and this step is applied at each symbol detection.} The slicing is extremely important in order to provide a proper interference cancellation and attain fully detector performance. Hence, if we assume that the estimated symbol in a determined layer is correct, the further symbols can be detected as if there were no previous layers, in a simple equivalent system. However, if an error occur on the first layers, it will propagate until the end of the algorithm, resulting in performance deterioration.

\subsection{{Ordered Successive Interference Cancellation (OSIC)}}
Improvements related to the BER performance of SIC can be attained through a suitable ordering scheme \cite{WubbenQR}, preventing error propagation during interference cancellation computation. The ordering criteria {have a focus to minimizing} the columns norm of $\mb{Q}$, {which cause the detection process to start from the highest normalized power symbol to the weakest one.}

The sorted decomposition can be expressed as:
\begin{equation}
	\mb{H}\mb{\Pi } = \mb{QR}
\end{equation}
where $\mb{\Pi}$ is a permutation matrix, {which allows symbols reordering after executing} the SIC detection. It is important to notice that the detection proceeds as a conventional SIC, {which is depicted by equation} \eqref{SIC}. {The only difference lay on the final step, where, by the end of the detection scheme, the reordering step is followed by multiplying the detected symbols vector with the permutation matrix.}

{The sorted QR decomposition (SQRD)} is {formalized} at the pseudocode in Algorithm \ref{alg:SQRD}. 
The main difference between this algorithm and the conventional QR decomposition, lay at the lines 2 and 3 of the algorithm \ref{alg:SQRD}. {Also, these lines do not carry out high complexity operation, causing the ordering complexity to be essentially negligible. For the rest of this work, this decomposition plus the detection scheme will be referred as ordered successive interference cancellation (OSIC).}

\begin{algorithm}
	\caption{Sorted QR decomposition \cite{Kobayashi_OSIC} 
		\label{alg:SQRD}}
	\begin{algorithmic}[1]
		\Require{$\mb{Q} = \mb{H}$, $\mb{R} = \mb{0}$, $\mb{\Pi} = \mb{I}_{n_T}$}
		\Ensure{$\mb{Q},\mb{R}$}
		\For{$i = 1 \textrm{ to } n_T$}
		\State	$k =  \displaystyle \argmin_{ j = i \textrm{ to } n_T}  \norm{\mb{q}_j}^2$
		\State Exchange columns $i$ and $k$ in $\mb{Q}$, $\mb{R}$ and $\mb{\Pi}$ 
		\State $r_{ii} = \abs{\mb{q}_i}$ 
		\State $\mb{q}_i = \mb{q}_i/r_{ii}$
		\For{$j = i + 1 : n_T $}
		\State $r_{ij} = \mb{q}^H_i\mb{q}_j$
		\State $\mb{q}_j = \mb{q}_j - r_{ij}\mb{q}_i$
		\EndFor
		\EndFor
	\end{algorithmic}
\end{algorithm}

Recently, Kobayashi\cite{Kobayashi_OSIC} have proved that the sorted QR decomposition based on the Gram-Schimid method is unable to operate satisfactorily at high SNR regime. It was also showed that SQRD algorithm based on Gram-Schimidt's was incapable to promote the orthonormalization of matrix channel $\mb{H}$ when the channel is highly correlated, which can make the $\mb{Q}$ matrix do not achieve the orthogonality and failing the OSIC requirements. Hence, the authors proposed a change in the norm update in the classic algorithm and numerically prove the stabilization of the OSIC in high SNR regime. Finally, the Algorithm \ref{alg:SQRD} is the modified version that can achieve better BER performance in the high SNR regime, while the same performance of the classic algorithm was held in low and medium SNR regions. 

\subsection{Lattice Reduction (LR) aided MIMO Detector}\label{sec:LR}
As already mentioned, if the channel matrix has an strongly spacial correlation characteristic or even a strong line-of-sight (LOS) component, the channel matrix become ill conditioned; which disrupts the detection process and mainly deteriorates the MIMO system performance. An ill conditioned matrix causes a narrowing on the symbol decision regions, which makes the the detection more vulnerable to even the smallest amount of noise. Hence, to circumvent this problem, we aim to turn the channel matrix as near-orthogonal as possible, looking for improve the MIMO detection process with a manageable complexity increase.

The {LR can be efficiently carried out} through the LLL algorithm, which was proposed by Lenstra-Lenstra-Lovaz in \cite{LLL}. However, for this entire work, is recommended the usage of the Algorithm \ref{alg:LR} to ensure the complex LR, which is known to be more robust for MIMO detection, furthermore presents less computational complexity \cite{MaWei}. 
\begin{algorithm}
	\caption{The Complex LLL Algorithm \cite{MaWei} (Using MATLAB Notation) \label{alg:LR}}
	\begin{algorithmic}[1]
		\Require{$\mb{H}$}. \qquad \Ensure{$\tilde{\mb{Q}},\tilde{\mb{R}},\mb{T}$}
		\State $\delta = 0.75$
		\State $m = $ columns number of $\mb{H}$
		\State {$\mb{T} = \eye_m$}
		\State $\sbrac{\widetilde{\mb{Q}},\widetilde{\mb{R}}} = \textrm{QR}(\mb{H})$ 	
		\State $k = 2$
		\While{$k\leq m$}
		\For{$n = k -1 $ to $1$}
		\State $u = \round{\dfrac{\widetilde{\mb{R}}\rbrac{n,k}}{\widetilde{\mb{R}}\rbrac{n,n}}}$
		\If{$u \neq 0$}
		\State $\widetilde{\mb{R}}\rbrac{1:n,k} = \widetilde{\mb{R}}\rbrac{1:n,k} - u \widetilde{\mb{R}}\rbrac{1:n,n}$
		\State $\mb{T}\rbrac{:,k} = \mb{T}\rbrac{:,k} - u \mb{T}\rbrac{:,n}$
		\EndIf
		\EndFor
		\If{$\delta \abs{\widetilde{\mb{R}}\rbrac{k-1,k-1}}^2 > \abs{\widetilde{\mb{R}}\rbrac{k,k}}^2 + \abs{\widetilde{\mb{R}}\rbrac{k-1,k}}^2$}
		\State {Swap the $(k-1)$-th and $k$-th columns in $\widetilde{\mb{R}}$ and $\mb{T}$}
		\State $\alpha = \frac{\widetilde{\mb{R}}\rbrac{k-1,k-1}}{\norm{\widetilde{\mb{R}}\rbrac{k-1:k,k-1}}_2}$ \quad and 
\quad $\beta = \frac{\widetilde{\mb{R}}\rbrac{k,k-1}}{\norm{\widetilde{\mb{R}}\rbrac{k-1:k,k-1}}_2}$
		\State $\Theta = \begin{bmatrix}
		\alpha^{\star} & \beta \\ -\beta & \alpha
		\end{bmatrix}$		
		\State $\widetilde{\mb{R}}\rbrac{k-1:k,k-1:m} = \Theta\widetilde{\mb{R}}\rbrac{k-1:k,k-1:m}$
		\State $\widetilde{\mb{Q}}\rbrac{:,k-1:k} = \widetilde{\mb{Q}}\rbrac{:,k-1:k}\Theta^H$
		\State $k = \max(k - 1,2);$
		\Else
		\State $k = k + 1$		
		\EndIf		
		\EndWhile
	\end{algorithmic}
\end{algorithm}

Basically, for detection purposes, the LLL algorithm {decomposes} the MIMO channel into a new base in a reduced domain:
\begin{equation}
	\widetilde{\mb{H}} = \mb{HT},
\end{equation}
where $\widetilde{\mb{H}}$ {is the reduced basis, offering improved properties regarding near-orthogonality when compared with the former} $\mb{H}$, while $\mb{T}$ is a unimodular matrix with {two properties:} $\det(\abs{\mb{T}}) = \pm 1$, and $\mb{T} \in \cbrac{\mathbb{Z}^{n_R \times n_T} + j\mathbb{Z}^{n_R \times n_T}}$. 

The new matrix $\widetilde{\mb{H}}$ {have better numerical conditioning features, essentially corresponded by larger decision regions allowing reductions at the linear detectors noise enhancement,} which {consequently} allows performance gain in the signal detection. The idea behind the LR-aided MIMO detection is to detect the symbols at the LR domain,{ so it's preferable to rework} the MIMO transmit equation in the LR domain {as follows}:
\begin{equation}
\arraycolsep = 1.4pt
\begin{array}{rcl}
\mb{x} &=& \mb{Hs} + \mb{n} \\
& = & \rbrac{\mb{H}\mb{T}}\rbrac{\mb{T}^{-1}\mb{s}} + \mb{n}\\
& = & \widetilde{\mb{H}}\mb{z}+ \mb{n}
\end{array}
\end{equation}

{Applying the reworked} system model, {the detection scheme under LR domain can be performed by any linear MIMO detection technique, such as ZF and MMSE}, optionally combined with the SIC or OSIC techniques. However, is extremely important to properly quantize the symbols in the reduced domain, this is performed through:
\begin{equation} \label{LRquant}
\widehat{\mb{z}} = \round{\dfrac{\widetilde{\mb{z}} - \beta'\mb{T}^{-1}{\ones_{n_T \times 1} }}{2}} + \beta'\mb{T}^{-1}{\ones_{n_T \times 1}}	
\end{equation}
where $\round{\cdot}$ {represents} the round operator, ${\mb{1}_{n_T \times 1}}$ {is an all ones column vector}, {$\widetilde{\mb{z}}$ is the estimated symbols after a MIMO detection strategy} and $\beta'$ is a constant {controlled} by the modulation order \cite{Milford}. For transmissions schemes that uses M-QAM modulation, we set $\beta' = \rbrac{1 + i}$ and for binary phase shift keying (BPSK) modulation we set $\beta' = 1$.

\subsection{LR aided Linear Equalization}
When linear detectors are considered, the equalization in the LR domain can be done in the exact same way as in sections \ref{sec:ZF} and \ref{sec:MMSE}, the only difference occurs in the quantization. Thus, for the ZF aided LR case, the solution is given as follows:
\begin{equation}\label{LRZF}
\arraycolsep = 1.4pt
\begin{array}{rcl}
\widetilde{\mb{z}} & = &\widetilde{\mb{H}}^{\dagger} {\mb{x}}\\
& = &\mb{z}+ \widetilde{\mb{H}}^{\dagger}\mb{n}.\\
\end{array}
\end{equation}
On the other hand, for the MMSE it is recommended the usage of the extended matrix due to its better performance. \cite{Wubben}. Thus, the LLL will be executed over the extended channel matrix, i.e,
\begin{equation}
\underline{\widetilde{\mb{H}}} = \underline{\mb{H}}\underline{\mb{T}}.
\end{equation}
{where $\underline{\mb{T}}$ is the unimodular matrix, {\it i.e.},  $\underline{\mb{T}}$ contains only integer entries and the determinant is ${\rm det}\left(\underline{\mb{T}}\right) = \pm 1$}.

Then, the MMSE solution in the LR domain is given as:
\begin{equation}
\widetilde{\mb{z}} = \underline{\widetilde{\mb{H}}}^{\dagger}{\underline{\mb{x}}}
\end{equation}

According to \cite{Wubben,WubbenQR}, {the noise term still contaminate the symbols at the LR domain and a convenient decision scheme must be executed at the symbols in this domain.}. As the LR operations consists in scaling and shifting the lattice points, it is necessary to include a re-scaling and re-shifting operations, that is given by the LR quantization in eq \eqref{LRquant}

Finally, the last step of the LR-aided MIMO detection consists in convert the estimated symbol vector from the LR domain {to the original constellation set}:
\begin{equation}
\widehat{\mb{s}} = \mb{T\widehat{\mb{z}}.}
\end{equation}	

\section{Performance Analysis}\label{sec:perform}
Next, the simulated BER performance of the previously discussed MIMO detectors have been compared. The system analysis is performed as a data transmission from the BS, with $n_T$ antennas to a MT equipped with $n_R$ antennas, i.e, the downlink scenario. In order to have a rightful comparison among different MIMO transmission set-ups, even when particular modulation order and number of antennas are applied, all performances will be examined under a normalized SNR in terms of {bit energy ($E_b$)}, as:
\begin{equation}
\dfrac{E_b}{N_0} = \dfrac{ SNR}{\log_2 M},
\end{equation}
where $M$ is the constellation order and $N_0=\frac{\sigma_n^2}{B}$ is noise power spectral density, with $B$ been the MIMO system bandwidth.
Besides, the transmit power constraint must be adopted, with power equally distributed among the $n_T$ antennas. 

Firstly, we consider an ULA distribution on the transmit and receive antennas, which results in the spatial correlation modeled in Section \ref{Corr}. We have considered three different arrangements of modulation order and number of antennas (modulation; $n_T\times n_R$) as evaluation standard for the MIMO detectors performance that do not generate prohibitive computational effort for the SD detector:
\begin{multicols}{3}
\begin{itemize}
\item[a)] ($16$-QAM; $8 \times 8$); \item[b)]  ($64$-QAM; $4 \times 4$); \item[c)] ($4$-QAM; $20 \times 20$).
\end{itemize}
\end{multicols}\vspace{-5pt}

We also consider an UPA distribution for both transmit and receive antennas. In this case, as the structure is considered to work within massive MIMO systems, it was considered structures with high number of antennas. The studied arrangement are listed bellow: 
\begin{multicols}{2}
\begin{itemize}
\item[a)] ($16$-QAM; $8 \times 8$); \item[b)] ($4$-QAM; $64 \times 64$) (massive-MIMO);
\end{itemize}
\end{multicols}\vspace{-5pt}
Thus, three antenna-correlation scenarios has been applied: $\rho = 0$,\,\, $0.5$ and $0.9$, which represents respectively no correlation, medium and strong correlation among antenna elements. Finally, in order of simplicity we have considered perfect knowledge of the channel gains in the receiver side, which means, the channel content $\mb{H}$ is available at the receiver, but unknown at the transmitter side.

\begin{figure}[!htbp]
	\begin{center}
		\includegraphics[width=0.76\textwidth]{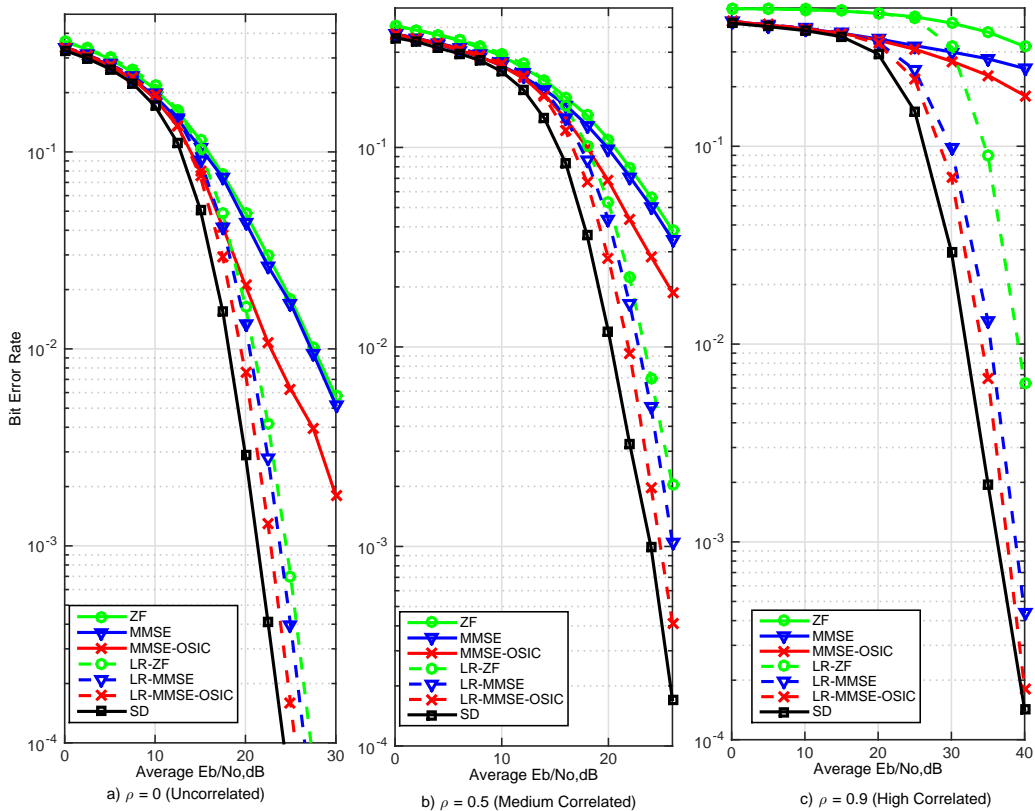}
	\vspace{-3mm}
		\caption{ BER for the first arrangement ($64-QAM; \, 4 \times 4$)}
		\label{fig:64QAM_4x4}
	\end{center}
\end{figure}

Figure \ref{fig:64QAM_4x4} {illustrates the BER performance in a MIMO system configured with $64$-QAM modulation and $4 \times 4$ antennas format.} We begin the analysis at low SNR regime, where all the analyzed MIMO detectors provide very similar performance. However, it is important to notice that the SD and the LR based detector can achieve full diversity, which means, in high SNR regime, where the SNR is negligible, there is a drop of $2^{n_T}$ in the BER for every $3$ [dB] increase in the SNR. As the system is based on a small antenna array, the LR-aided detectors have an excellent performance showing a narrow gap in comparison with the SD, and also their BER curve remains parallel to the SD one, which implies in same diversity order.
\vspace{-3mm}
	
\vspace{-6mm}
	
Both ZF and MMSE detectors have similar performances in high SNR regions, this statement is verified by equations \eqref{ZF} and \eqref{MMSE}, where the difference is that the noise statistics are considered at the MMSE equation. Furthermore, by the application {of} interference cancellation, lattice-reduction techniques or the combination of both techniques leads to a great improvement in the MIMO detection performance, which is verified at Figure \ref{fig:64QAM_4x4}.

Regarding the performance impact due to antenna correlation, {is expedite} confirm that as the correlation index increase the BER performance {degenerates.} At high correlation scenario, the non-LR-aided detectors {require} very high SNR to operate in {suitable} BER levels. {This SNR demand for highly correlated scenarios directly impact in the energy efficiency, leading to undesirable rates.} In fact, {exclusively} SD and LR-based MIMO detectors {enables} a great transmission energy efficiency and full diversity under high antenna correlation, which results in great BER performance, as seen in Figure \ref{fig:64QAM_4x4}(c).

\begin{figure}[!htb]
\begin{center}
\includegraphics[width=.75\textwidth]{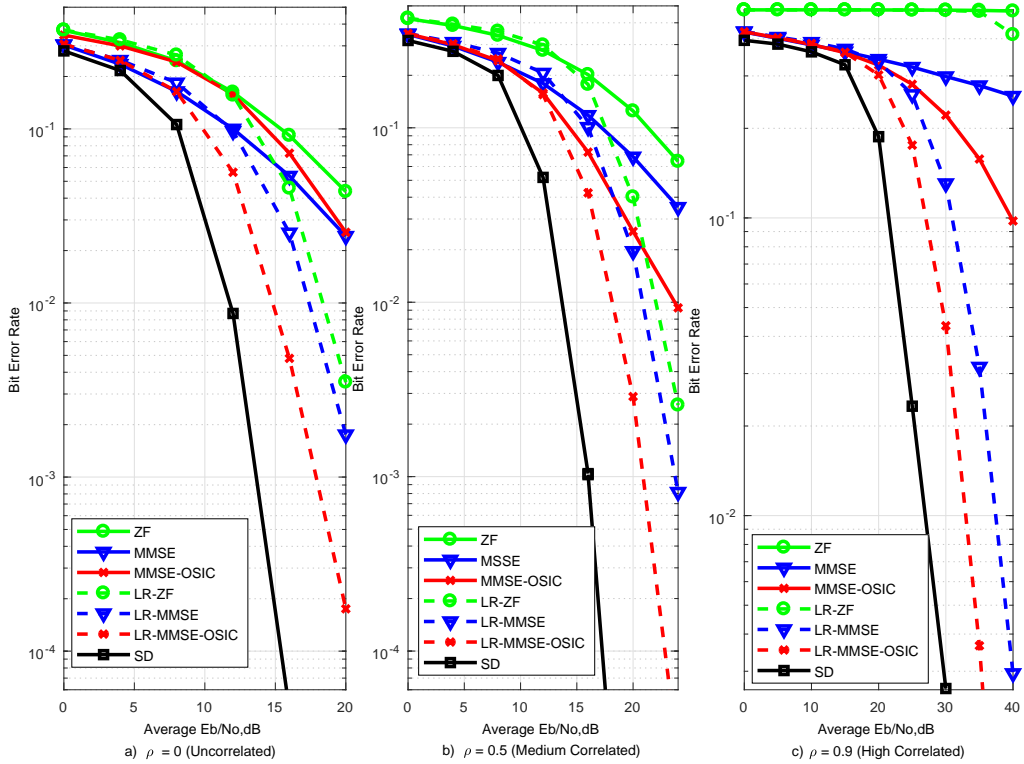}
\caption{BER for the second arrangement ($16-QAM; \, 8 \times 8$)}
\label{fig:16QAM_8x8}
\end{center}
\end{figure}

Increasing the number of antennas, i.e. the ($16$-QAM; $8 \times 8$) case, will make the BER gap between the SD and the other MIMO detectors also to be increased, which is noticed in Figure \ref{fig:16QAM_8x8}. With this arrangement, differences in BER performance {are evident}; the most notable performance is achieved when the MMSE detector is combined with both LR and OSIC techniques, which was the closest to the optimal.
\vspace{-5mm}
	
\vspace{-5mm}

\begin{figure}[!htbp]
\begin{center}
\includegraphics[width=.75\textwidth]{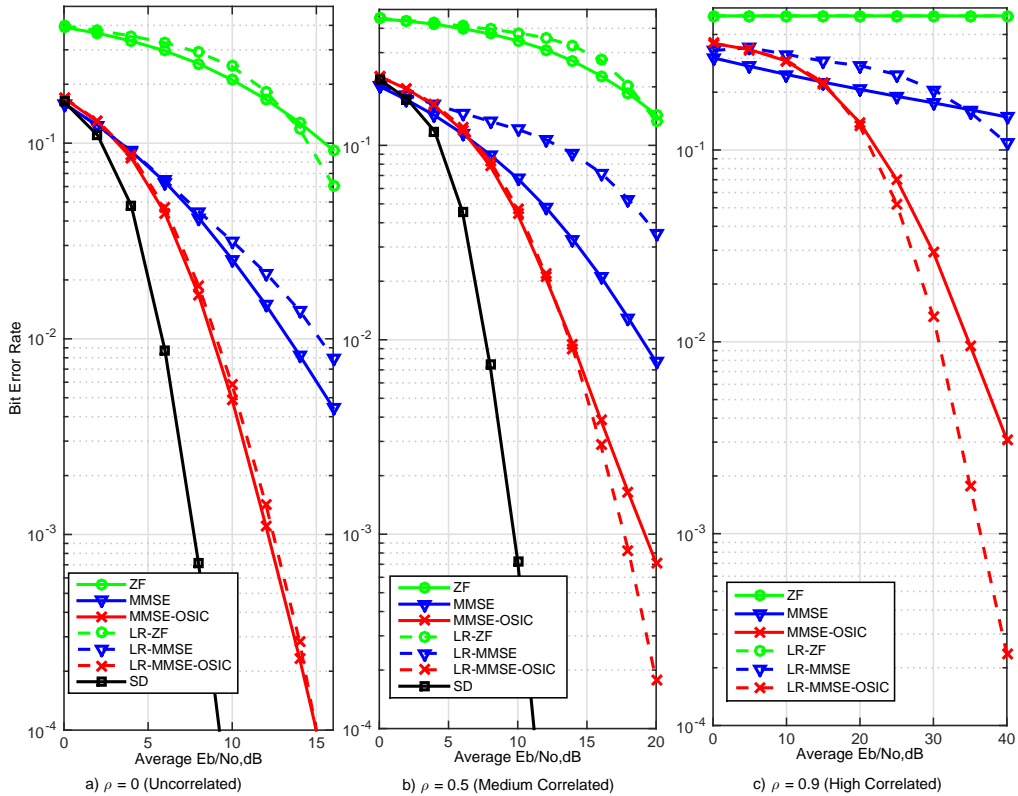}
\caption{BER for the third arrangement ($4-QAM \,\, 20 \times 20$ antennas)}
\label{fig:4QAM_20x20}
\end{center}
\end{figure}

It is important to notice that in large antenna arrays, such as Figure \ref{fig:4QAM_20x20} with $n_R = n_T = 20$ antennas, the BER performance behaves different in each detection case. The first point is related to the ZF detector which becomes inefficient at low and medium space correlation scenarios, requiring high SNR regime to achieve reasonable BER performance. {At high correlated scenarios, i.e $\rho = 0.9$, the ZF detector completely fails in decoupling the inter-antenna interference, also the MMSE detector loses diversity}, while the LR-MMSE suffers from great BER performance degradation, particularly in high SNR regime. Furthermore, despite it extremely superior performance, under high spacial correlated channels the SD MIMO detector has showed an extremely exceeding computational complexity due to the vast number of branches that the SD algorithm needs to visit in order to detect the symbols in this configuration.

\subsection{Spatial UPA $\times$ ULA Correlated Channels}
\begin{figure}[!htbp]
\centering
\includegraphics[width=0.7\textwidth]{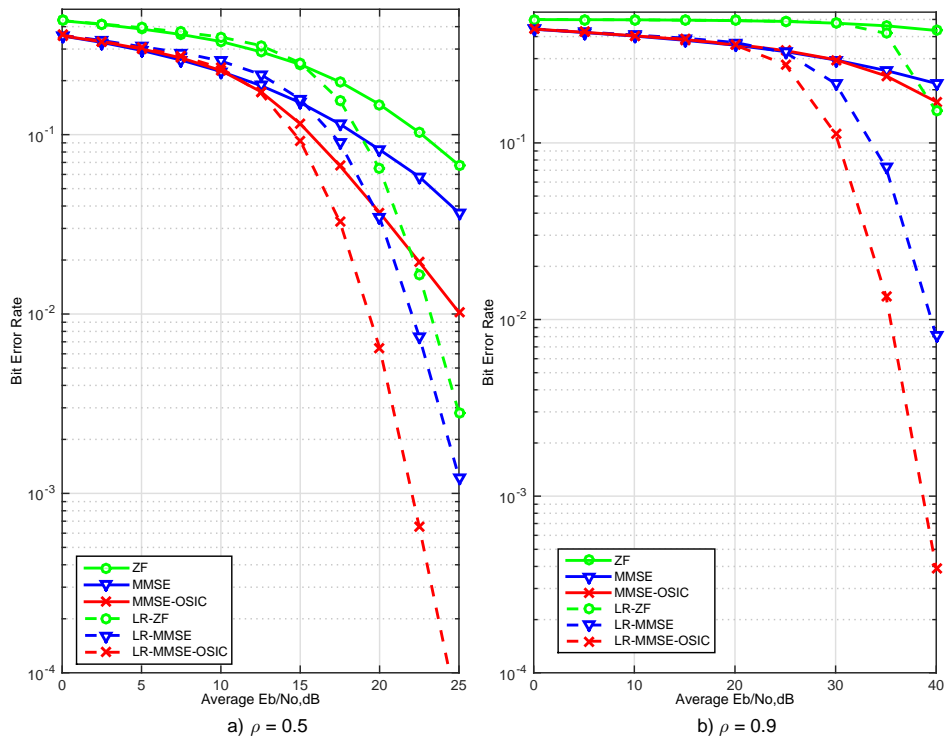}\\
\vspace{-1mm}
\caption{BER for the first UPA arrangement ($16$-QAM; $8 \times 8$), with medium ($\rho = 0.5$) and very high ($\rho = 0.9$) spatial correlation.}
\label{fig:16QAM_8x8_URA}
\end{figure}

Figure \ref{fig:16QAM_8x8_URA} depicts the BER performance for a $16$-QAM MIMO with $n_R = n_T = 8$ with UPA antenna array (from Fig. \ref{fig:UPAcorr}) deployment and correlation index $\rho = 0.5$. Note that with the UPA array geometry applied the correlation effect becomes more severe due to the inner geometrical problem which is related to the antenna elements position. Notice that in uncorrelated channel scenarios the BER performance achieved for both ULA and UPA arrays will be the same for any system configuration, {due to} the Toeplitz structure of the correlation channel matrix. Otherwise, in correlated channel scenarios, performance losses will be expected for both UPA and ULA arrays, but with {higher} losses in the UPA structure, due to the cross-distances within antenna elements at both $x$ and $y$-axes of the Euclidean plane. {Such arrangement} leads to higher interference in the received signal, leading to noise enhancement and consequently BER performance losses. 

\vspace{-5mm}

\vspace{-5mm}

\begin{figure}[!htb]
\centering
\includegraphics[width=0.6\textwidth]{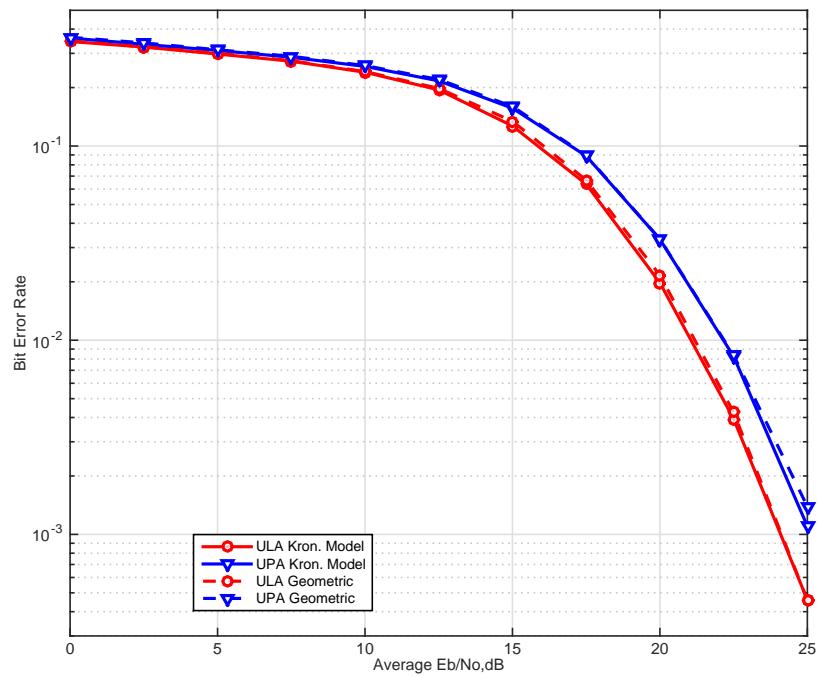}
\vspace{-1mm}
\caption{BER $\times E_b/N_0$ for the LR-MMSE detector with ($16$-QAM; $8 \times 8$) arrangement, correlation $\rho = 0.5$ with ULA and UPA arrays deployed at both Kronecker's approximation and Geometrical Model. }
\label{fig:16QAM_8x8_URA-ULA_LR-MMSE}
\end{figure}

Figure \ref{fig:16QAM_8x8_URA-ULA_LR-MMSE} depicts the BER performance with the same previous configurations of modulation order and system size. The difference is that only the LR-MMSE detector is considered, in order to evaluate the performance gap between the system with different array structures. The detector choice {was} made based on the LR-MMSE capacity to {maintain} the diversity at high SNR regions, and also because the characteristic of being able to deal with correlated channels {while} keeping a great performance. It is expedite conclude that, at medium and high correlation index there is a tiny performance gap between the ULA and UPA, in the order of approximately 2dB, {which} is introduced due to the correlation matrix condition. {Specifically, as there is less interference} coming from the neighbor antennas in the ULA correlated channel, a slightly better BER performance is {observed.} 

Another interesting result depicted in \ref{fig:16QAM_8x8_URA-ULA_LR-MMSE} is the approximation between the Kronecker and the geometrical correlation models. The BER comparison were obtained for moderate correlation index, $\rho = 0.5$, which is straightforward calculated under the Kronecker model and the Geometrical-based correlation model following the expressions \eqref{eq:geoULA} and \eqref{eq:geoUPA}, respectively, for ULA and UPA antenna arrangements. The $\mb{R}_{\rm ula}$ were simulated using the distance between elements $d = 0.5\lambda$ and the other variables are set to: $\theta = 3\pi/8, \; \xi = \pi/8$.  For the UPA arrangement, the $\mb{R}_{\rm h}$ were simulated using $d_1 = d_2 = 0.5\lambda$ and the other parameters were set to: $\theta = 3\pi/8, \; \phi = \pi/3 \;\xi = \pi/8 $ and $\delta = \pi/6$ which represents a slightly greater angular spread when compared to the one used in \cite{Ying}, because as larger the angular spread, lower is the correlation index and we needed to find a moderated correlated case, lower than the used by the refereed authors.

\subsubsection{UPA Correlated Channels under Large-Scale Configuration}
\begin{figure}[!htbp]
\centering
\hspace{-10mm}\includegraphics[width=0.73\textwidth]{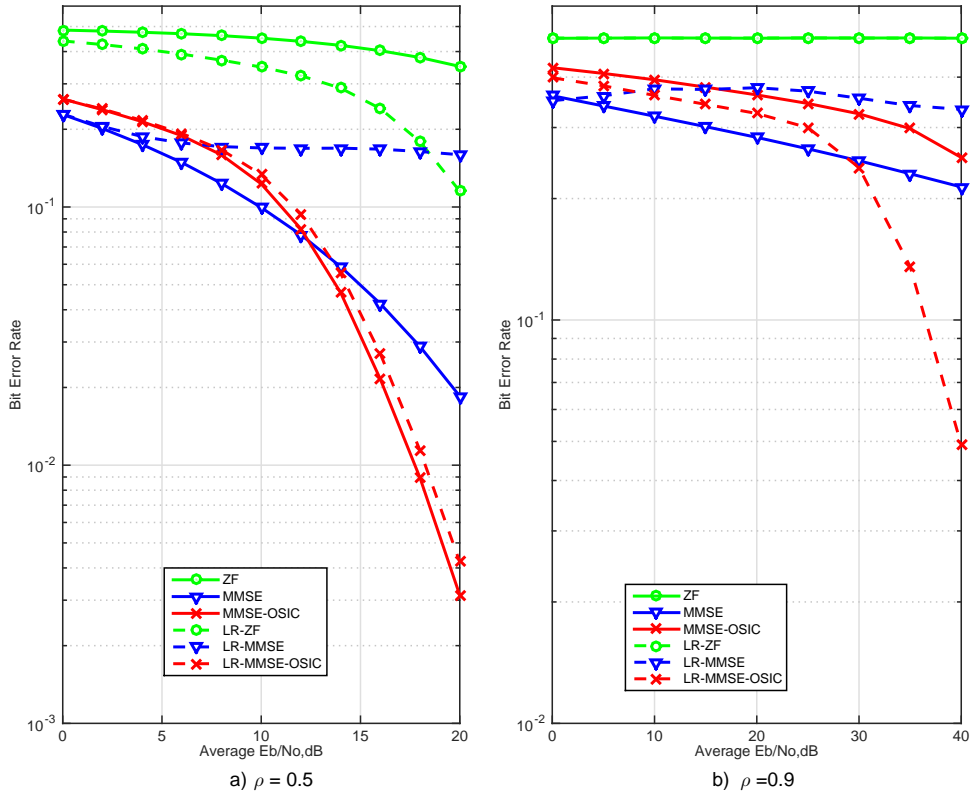}\\
\vspace{-2mm}
\caption{BER for the second UPA arrangement ($4$-QAM $64; \times 64$)}
\label{fig:4QAM_64x64_URA}
\end{figure}

The UPA structure is proposed when large antennas array structures are deployed at the base station; following this perspective, Figure \ref{fig:4QAM_64x64_URA} depicts the BER performance for a $4$-QAM $64 \times 64$ antennas systems. In this arrangement the SD detector performance is not depicted due to its impractical complexity over high number of antennas. With correlation index $\rho = 0.5$, all detectors tend to show greater degradation especially the LR-aided ones. This behavior can be explained due to the high size on the channel matrix, which makes more difficult to find a new orthogonal basis; as expected for high sized channel correlated matrix, the MMSE-OSIC detector presents a very similar performance of {its} LR-aided version. Furthermore, when the correlation index is incremented to a high correlated scenario, $\rho = 0.9$, all detectors suffer large {diversity losses}, except for the LR-MMSE-OSIC, which, despite the high-scale scenario, compared to other detectors, is still able to achieve greater diversity under high SNR regime.

\vspace{-5mm}
\subsection{Array Gain Impact on the Performance}
The last analysis in this section is related to the array factor (AF), or array gain, {which directly impacts the transmission gain}, that eventually will impact over the SNR. As seen in section \ref{Corr} the array factor for both structures will vary accordingly to the number of antennas and the spacing between them. It is also verified in that section that UPA structures are able to provide much more gain over the transmit direction regarding the uniform linear array. The impact of this characteristic is completely related to the BER performance, because the greater the normalized power loss, the worse will become the BER performance. In this perspective, it is important to emphasize that all previous BER performance results for ULA and UPA were conceived considering an array gain of $0$dB, which implies that the beam gain from the BS is directed to the MT in a point-to-point MIMO link configuration. In terms of elevation and azimuth angles $\theta = 0^{\circ}$ and $\phi = 0^{\circ}$, {and in this case, only the correlation effect will impact the BER performance.} 

\begin{figure}[!htb]
	\begin{center}
		\includegraphics[width=0.78\textwidth]{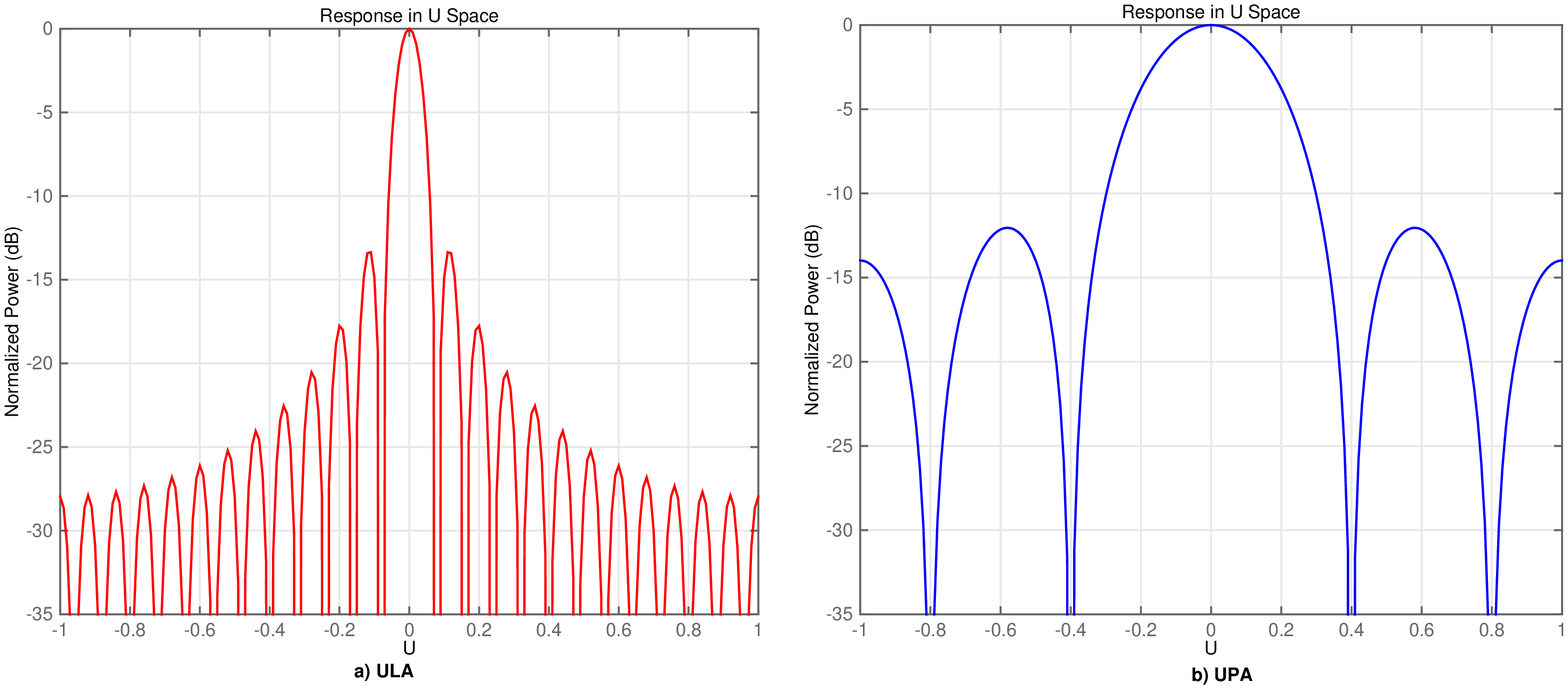}
		\vspace{-2mm}
		\caption{UV response under azimuth angle $\phi = 0^{\circ}$ condition considering: {\bf a)} $25$ antennas ULA; {\bf b)}  $5 \times 5$ UPA with $0.5\lambda$ element-spacing.}
		\label{fig:UVsec}
	\end{center}
	\vspace{-5mm}
\end{figure}

A comparative analysis on the array gain for both array structures were made based on a $5 \times 5$ UPA and a $25$ element ULA with $0.5\lambda$ element-spacing. The array gain comparison is provided in Table \ref{tab:AG}. To do such analysis, it is necessary to compare the UV response for both array structures. Figure \ref{fig:UVsec} provide the UV response for both array structures in the azimuth cut condition, which means the azimuth angle {is} $\phi = 0^{\circ}$, leaving only the normalized power response for elevation, $\theta$, variations. Remembering that the $x$-axes follows the orthogonal projection given by equation \eqref{eq:project}, {adopting} $\phi = 0$, {$\theta = \arcsin(u)$}.

\begin{table}[!htp]
	\caption{UPA and ULA array gain over various elevation angles $\theta$ under azimuth angle $\phi = 0^{\circ}$ condition} 
	\centering
	\begin{tabular}{cc|c|c} \hline
		\multicolumn{2}{c|}{Elevation angle}&{ULA} & \multicolumn{1}{c}{UPA}  \\ \hline 	
		{$u$}&$\theta$ [$^{\circ}$] & Array Gain (dB) & Array Gain (dB)\\
		\hline \hline 
		$0$	 & $0$	  & $0$ & $0$  \\
		$0.12$	 & $6.9$  & $-13.41$  & $-1.27$ \\
		$0.2$   & $11.5$ & $-17.8$   & $-3.8$ \\
		$0.44$  & $26$   & $-24.05$  & $-20.2$ \\				
		$0.6$   & $37$   & $-26.12$  & $-12.4$ \\ 
		$0.86$  & $60$   & $-30$     & $-20$ \\\hline
	\end{tabular} 
	\label{tab:AG}
\end{table}

The array gain feature is directly related to the normalized power distribution, and each array structure provide its own power order. Analyzing the results in Table \ref{tab:AG}, it shows that a UPA structure provide greater gains, independently of the MT position. The only exception occurs when $\theta = 0^{\circ}$, because the BS will provide the same beam gain on the transmission for both ULA and UPA. In a MIMO point-to-point case, where the correlation effect is considered in both transmit and receiver side, the ULA structure will provide better performance {in cases where} the antenna beam pattern is focused on the MT. As the array gain {directly impacts} the SNR, and as consequence in the BER performance, cases where the antenna beam pattern is not focused directly on the MT, but in a region that has lower gain coverage, the UPA will provide better performances due to its higher gain lobe, especially under slightly deviations.

\section{Complexity Analysis }\label{sec:complexity}
The complexity analysis of MIMO detectors {is of} great importance, since all MT's should operate under strong signal processing and energy consumption limitations. With the combined analysis of BER performance and complexity it is possible to attain the best trade-off among the available detectors that comply with the system requirements.

With this objective in mind, this section presents a complexity comparison of those sub-optimum MIMO detectors. The complexity of each MIMO detector was measured in terms of flops (floating point operations), counting the total number of flops needed to perform the detection of a single transmitted symbol vector. For simplicity, we have considered the flop counting for complex operations, specifically, one flop was considered for summations and three flops for complex product \cite{WubbenQR}. Furthermore, the flop counting for matrix operations were based on \cite{Golub}, with the necessary modifications. Also, {the SQRD complexity} were {based} on \cite{WubbenQR}, also the complexity for the SD is {determined at} the study carried out in \cite{SD_complexity}. 

Through these methods, Table \ref{tab:flops} presents the complexity in terms of number of flops for each used matrix operations, including matrix multiplication, inversion, approximated LLL complexity and the QR decomposition, which are  procedures deployed in several MIMO detectors, specially those detectors treated herein,  Table \ref{tab:Complexity_MIMO}, where, $n = n_R = n_T$ and $M$ are the number of antennas and the M-QAM order of modulation, respectively.
\begin{table}[!htp]
	\caption{Number of flops for each operation/procedure } 
	\centering
	\begin{tabular}{l c} \hline
		Operation & Number of flops  \\ \hline 		
		$\mb{C}_{n\times n}= \mb{A}_{n \times n} \times \mb{B}_{n \times n}$ &  $2n^3 $\\
		$\mb{y}_{n\times p}= \mb{A}_{n \times n} \times \mb{x}_{n \times p}$ & {$2n^2p$} \\
		$\mb{C}_{n\times n}= \mb{A}_{n \times n} + \mb{B}_{n \times n}$ &  $n^2 $  \\
		$f_\textsc{lll}\rbrac{n,\rho} $ \cite{Koba} & $ \approx (ae^{b\rho} + c)n^3 $\\
		{SQRD \cite{WubbenQR}} & {${16n^3}/{3} + {7n^2}/{3} + {25n}/{6}$}\\				
		$\mb{C}_{n\times n}= \mb{A}^{-1}_{n \times n}$ & ${2n^3}/{3}$\\ \hline
	\end{tabular} 
	\label{tab:flops}
\end{table}

The LR aided detectors have presented decent BER performance, but in complexity meaning, it may offer an increasing one at determined scenarios. Through the simulations, it was observed that the complexity of the LLL algorithm does not only depend on the matrix size, but also on the correlation index. Naturally, the relation between matrix dimensions and complexity is straightforward due to increasingly operations number. On the other hand, as the matrix become more correlated it nears a quasi-singular matrix condition, making it difficult for the LLL algorithm to reach an orthogonal basis, leading to an increase in computational complexity.

The exact LLL complexity cannot be easily evaluated due to all the variable dependencies. However, it is known that {a} good approximation for the LLL complexity can be evaluated as a $\mathcal{O}(n^3 \log n)$ order \cite{LLL_Comp}. Aiming to provide a better expression {that} represent the LLL complexity, in \cite{Koba} a numerical experiment was conducted to determine, through the better surface fitting, the LLL complexity dependency w.r.t. the antenna correlation index and array dimension. With such experiment {the most similar surface fitting the LLL complexity, were given by:}
\begin{equation}\label{eq:LLLcompapprox}
	f_\textsc{lll}\rbrac{n,\rho} \approx \rbrac{ae^{b\rho} + c}n^3
\end{equation} 
with $a = 5.018\times 10^{-4}$, $b = 13.48$ and $c = 8.396$. Finally, for the LR-aided MIMO detectors, the necessary flop counting approximation for the LLL procedure given in \eqref{eq:LLLcompapprox} was included in the total complexity calculation.

\begin{figure}[!htb]
	\begin{center}
\includegraphics[width=0.65\textwidth]{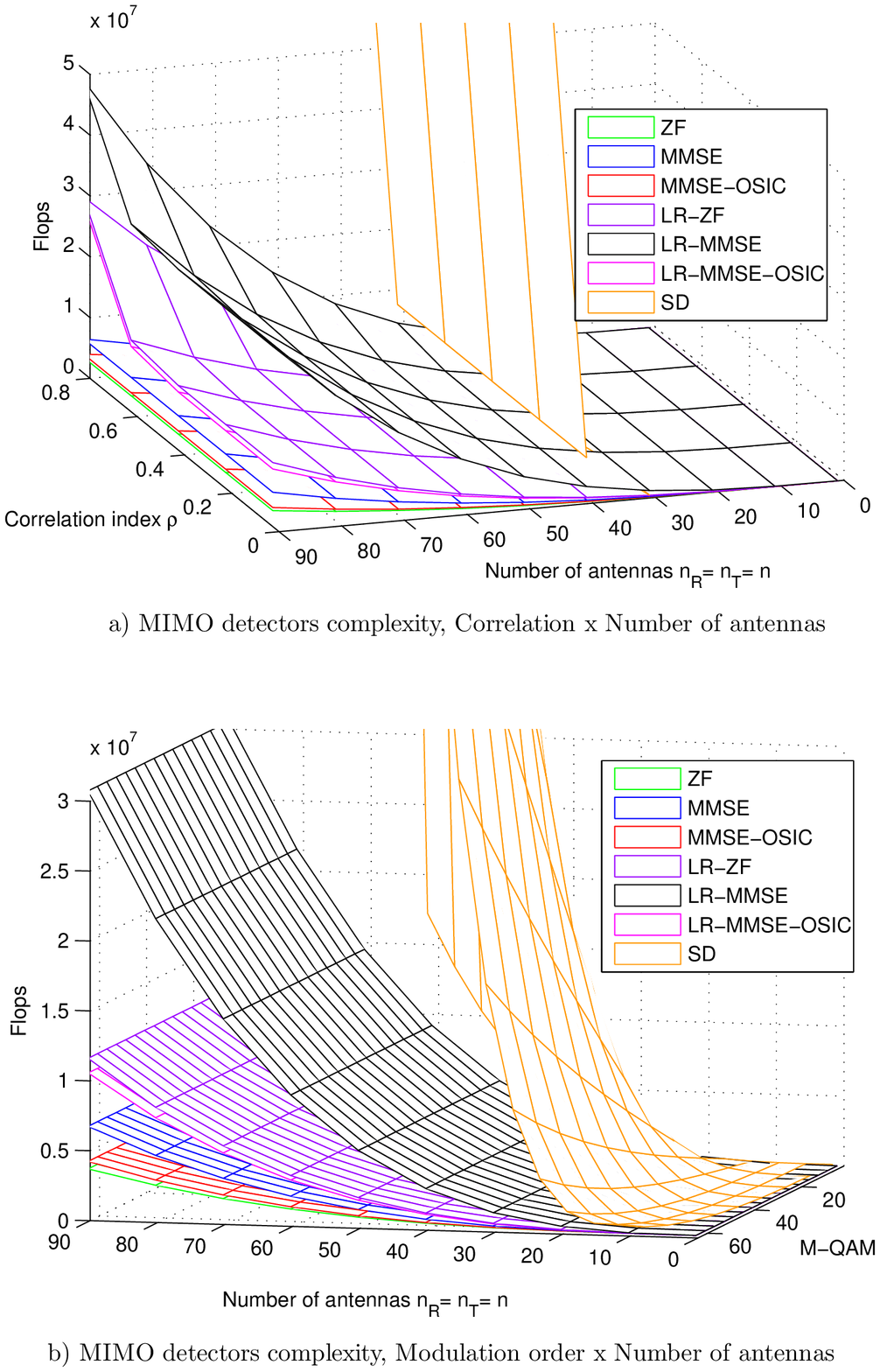}
\vspace{-4mm}
\caption{MIMO detector complexity $E_b/N_0 = 20[dB]$}
\label{fig:Comp_MIMO}
	\end{center}
\end{figure}

A complexity evaluation of the previous analyzed MIMO detectors and the various combinations possibilities have been made. Table \ref{tab:Complexity_MIMO} summarizes the overall complexity for the most relevant combinations of sub-optimum MIMO detectors covered in this work. Notice that the ML complexity grows exponentially and it becomes prohibitive when the product number of antennas by modulation order ($n\cdot M$) increases, {which is the case of any practical MIMO case of interest (including small-medium $n\cdot M$ values) \cite{Koba}.} Moreover, the SD complexity is not trivial to obtain; as demonstrated in \cite{SD_complexity} the SD complexity always present an exponential asymptotic behavior in low SNR and/or large $n\cdot M$ scenarios. This occurs because the algorithm needs to ensure certain probability to find a point inside the sphere, then if the problem size and/or noise power increase, the hyper-sphere radius grows and consequently the complexity. 

\begin{table}[!htp]
	\caption{MIMO Detectors Complexity} 
	\centering
\resizebox{0.55\textwidth}{!}{%
\begin{tabular}{l l} \hline
\bf MIMO Detector & \bf Total Complexity  \\ 
\hline\hline 		
{ZF \cite{RV}} & {\large ${14n^3}/{3}  + {2}n^2$}\\
{MMSE \cite{RV}} & {\large $26n^3/{3} + 4n^2$} \\
MMSE-OSIC & {\large $16n^3/3+ 13n^2 /3+ 25n/6$}  \\
LR-ZF & {\large $20n^3/3  + 10n^2 + 4n +f_\textsc{lll}\rbrac{n,\rho}$ }\\
LR-MMSE & {\large $32n^3 + 14n^2+ 3n + f_\textsc{lll}\rbrac{n,\rho}$} \\
LR-MMSE-OSIC & {\large $22n^2 /3+ 13n^2 /3+ 25n/6  +  f_\textsc{lll}\rbrac{n,\rho}$}  \\ \hline  
SD  & {\large $4n^3 + 7n^2 + n/2 + (2n + 2)\frac{M^{\gamma n} - 1 }{M -1}$,} \\
Ref.\cite{SD_complexity} & where $\gamma = {1/2{\sbrac{\frac{c^2(M^2 - 1)}{6{N_0} + 1}}}}^{-1}$ and $c^2 = \expect{\norm{\mb{h}_i}^2}$, $\forall i \in \sbrac{1,n}$ \\ 
\hline
{ML \cite{Koba}} &  {\large $M^{n}(4n^2 + 2n)$} \\
\hline
\end{tabular}
}
\label{tab:Complexity_MIMO}
\end{table}

{Regarding the ZF and MMSE detectors, their computational effort directly relates to Eq. \eqref{ZF} and \eqref{MMSE} respectively, which can be calculated through a matrix inversion, a matrix summation, and multiplications. The main difference between their complexity is that the MMSE requires two multiplications instead of one, and four multiplications, instead of two needed for the ZF algorithm.  The ZF and MMSE complexities comply with Ambrosio {\it et al} \cite{RV}.}

{When it comes to the OSIC-aided detectors, the primary source of computational effort it is the SQRD algorithm, which offers a cubic complexity order \cite{WubbenQR}. When the SQRD procedure is combined with the SIC algorithm, represented by Eq. \eqref{SIC}, the result is an increment in the quadratic order, due to the SIC computational method. The complexity for the MMSE-OSIC in Table \ref{tab:Complexity_MIMO} is corroborated by that found in \cite{Koba} and \cite{RV}.}

{Considering the LR aided detectors, the same procedure holds, but now with the addition of the LLL algorithm complexity. Following Algorithm \ref{alg:LR}, we notice that the LR is composed by a QR decomposition and the LLL, so the LR-ZF aided detector complexity can be determined from the equation, Eq. \eqref{LRZF}, in combination with the LLL algorithm and the matrix manipulations covered in section \ref{sec:LR}. Regarding the LR-MMSE the only difference is the usage of the extended matrix, which increases the computational effort by doubling the size of all operations. Finally, the LR-MMSE-OSIC is based on the addition of the SQRD algorithm, the LLL, and the SIC procedure, which reduce the complexity by eliminating a series of matrix multiplications. The complexity evaluation of the LR aided detectors relies upon \cite{Koba}.}

Figure \ref{fig:Comp_MIMO} depicts the computational complexity for the various MIMO detector studied in this work, divided in terms of flops as a function of: {\bf a}) normalized correlation index $\times$ number of antennas; {\bf b})  M-QAM order $\times$ number of antennas. From Table \ref{tab:Complexity_MIMO} and {Figure \ref{fig:Comp_MIMO}}, one can notice that the OSIC detectors are capable to offer much better complexity-performance tradeoffs when compared to the versions with the pseudo-inverse. This is caused by the fact that the SQRD leaves an upper triangular systems which demands lower complexities then the pseudo inverse. Also, ordered version is preferable over the simple SIC, {since, accordingly \cite{WubbenQR}, the first one requires $2n^2 - 2n$ flops in the overall complexity}, while providing considerable performance improvements. 

Moreover, Figure \ref{fig:Comp_MIMO}.b) {shows that the LR-aided detectors presents acceptable complexity under low or medium correlated channels; besides, {full diversity is maintained in these} scenarios, which turns this class of sub-optimal MIMO detectors one of the most promising in the context of this work.} The SD detector can result in great complexity saving only for cases where the systems is under high SNR regime with low number of antennas and modulation order, otherwise it becomes burden (increasingly high computational complex).

\section{Conclusions}\label{sec:concl}
The initial MIMO detection performance analyses carried out in this work were based on the independent identically distributed and perfect estimated channels; this opened the possibility to analyze the performance $\times$ complexity trade-off of such MIMO detectors operating under more realistic correlated channels. 

Lattice reduction technique has been proved to provide great BER performance improvements of linear sub-optimum MIMO detectors. The analysis of MIMO detectors under correlated channels indicates notable advantage in terms of reduction in BER degradation for the LR-aided MIMO detector due to the ability to deal with the near orthogonality of the channel matrix $\bf H$, besides full diversity achieved. The LR-MMSE-OSIC MIMO detector presented the smaller degradation in terms of BER performance, even under high correlated MIMO channels. Linear detectors aided by the combination of both LR and OSIC techniques can provide a near optimum performance in some cases; however the LR aided detectors tend to present increasingly complexity when high correlated scenarios are applied, which results by the orthonormalization difficult that the LLL algorithm shows when a near singular matrix is given as input. Therefore, the LR-MMSE-OSIC have achieved the best performance-complexity trade-off among the presented detectors.

When it comes to array structure and correlation effect, the ULA will always perform better when the antenna beam gain is focused on the MT, otherwise the UPA structure will provide greater BER performances, despite the correlation, due to its great inherent transmit power distribution pattern.

\section*{Acknowledgement}
This work was supported in part by the National Council for Scientific and Technological Development (CNPq) of Brazil under Grants 304066/2015-0, and in part by CAPES - Coordenação de Aperfeiçoamento de Pessoal de Nível Superior, Brazil (scholarship),  and by the Londrina State University - Paraná State Government (UEL).


\begin{thebibliography}{10}

\bibitem{wubbenLR}
Wubben Dirk, Seethaler Dominik, Jald{\'e}n Joakim, Matz Gerald. Lattice
  reduction.  {\it IEEE Signal Processing Magazine. }2011;28(3):70--91.

\bibitem{VB}
Wolniansky P.~W., Foschini G.~J., Golden G.~D., Valenzuela R.~A.. V-BLAST: an
  architecture for realizing very high data rates over the rich-scattering
  wireless channel.  In: :295-300; 1998.

\bibitem{Boccardi}
Boccardi F., Heath R.~W., Lozano A., Marzetta T.~L., Popovski P.. Five
  disruptive technology directions for 5G.  {\it IEEE Communications Magazine.
  }2014;52(2):74-80.

\bibitem{ScalingUp}
Rusek F., Persson D., Lau B.~K., et al. Scaling Up MIMO: Opportunities and
  Challenges with Very Large Arrays.  {\it IEEE Signal Processing Magazine.
  }2013;30(1):40-60.

\bibitem{Jalden2}
Jalden Joakim. Maximum Likelihood Detection for the Linear MIMO Channel.
\newblock PhD thesisRoyal Institute of Technology2004.

\bibitem{Barbero}
Barbero L.~G., Thompson J.~S.. Fixing the Complexity of the Sphere Decoder for
  MIMO Detection.  {\it IEEE Transactions on Wireless Communications.
  }2008;7(6):2131-2142.

\bibitem{Bohnke}
Bohnke R., Wubben D., Kuhn V., Kammeyer K.~D.. Reduced complexity MMSE
  detection for BLAST architectures.  In:  IEEE Global Telecommunications
  Conference, vol. 4: :2258-2262; 2003.

\bibitem{WubbenQR}
Wubben D., Bohnke R., Kuhn V., Kammeyer K.~D.. MMSE extension of V-BLAST based
  on sorted QR decomposition.  In:  VTC 2003-Fall -- IEEE 58th Vehicular
  Technology Conference, vol. 1: :508-512; 2003.

\bibitem{RV}
Valente Raul~Ambrozio, Marinello Jos{\'e}~Carlos, Abr{\~a}o Taufik. LR-aided
  MIMO detectors under correlated and imperfectly estimated channels.  {\it
  Wireless personal communications. }2014;77(1):173--196.

\bibitem{MaWei}
Ma~Xiaoli, Zhang Wei. Performance analysis for MIMO systems with
  lattice-reduction aided linear equalization.  {\it Communications, IEEE
  Transactions on. }2008;56(2):309--318.

\bibitem{Wubben}
Wubben D., Bohnke R., Kuuhn V., Kammeyer K.-D.. Near-maximum-likelihood
  detection of MIMO systems using MMSE-based lattice reduction.  In:
  Communications, 2004 IEEE International Conference on, vol. 2: :798-802;
  2004.

\bibitem{larsson2009mimo}
Larsson Erik~G. MIMO detection methods: How they work [lecture notes].  {\it
  IEEE Signal Processing Magazine. }2009;26(3):91--95.

\bibitem{Bai}
Bai Lin, Choi Jinho, Yu~Quan. {\it Low Complexity MIMO Receivers}.
\newblock Springer Publishing Company, Incorporated; 2014.

\bibitem{Koba}
Kobayashi Ricardo~Tadashi, Ciriaco Fernando, Abr{\~a}o Taufik. Efficient
  Near-Optimum Detectors for Large MIMO Systems Under Correlated Channels.
  {\it Wireless Personal Communications. }2015;83(2):1287--1311.

\bibitem{Cho:2010}
Cho Yong~Soo, Kim Jaekwon, Yang Won~Young, Kang Chung~G.. {\it MIMO-OFDM
  Wireless Communications with MATLAB}.
\newblock Wiley Publishing; 2010.

\bibitem{Goldsmith:2005}
Goldsmith Andrea. {\it Wireless Communications}.
\newblock New York, NY, USA: Cambridge University Press; 2005.

\bibitem{ULA_real}
Mandeep {J. S.}, Misran N., Abdullah H., How {Tan Chiy}. Patch array antenna
  serves satcom needs.  {\it Microwaves and RF. }2010;49(4).

\bibitem{van}
Van~Zelst A, Hammerschmidt JS. A single coefficient spatial correlation model
  for multiple-input multiple-output (MIMO) radio channels.  In: Proc. of URSI
  General Assembly:17--24; 2002.

\bibitem{Balanis}
Balanis Constantine~A.. {\it Antenna Theory: Analysis and Design}.
\newblock Wiley-Interscience; 2005.

\bibitem{UPA_real}
Gao X., Edfors O., Rusek F., Tufvesson F.. Linear Pre-Coding Performance in
  Measured Very-Large MIMO Channels.  In: 2011 IEEE Vehicular Technology
  Conference (VTC Fall):1-5; 2011.

\bibitem{URA_model}
Levin G., Loyka S.. On Capacity-Maximizing Angular Densities of Multipath in
  MIMO Channels.  In: 2010 IEEE 72nd Vehicular Technology Conference -
  Fall:1-5; 2010.

\bibitem{URA_2}
Li~J., Su~X., Zeng J., et al. Codebook Design for Uniform Rectangular Arrays of
  Massive Antennas.  In: 2013 IEEE 77th Vehicular Technology Conference (VTC
  Spring):1-5; 2013.

\bibitem{Ying}
Ying D., Vook F.~W., Thomas T.~A., Love D.~J., Ghosh A.. Kronecker product
  correlation model and limited feedback codebook design in a 3D channel model.
   In: 2014 IEEE International Conference on Communications (ICC):5865-5870;
  2014.

\bibitem{ZhaoWang}
Zhao Y., Wang X., Yang J., Zhao B.. Downlink Closed-Loop Training Sequence
  Design for Massive MIMO Systems with Uniform Planar Arrays.  In: 2016 IEEE
  83rd Vehicular Technology Conference (VTC Spring):1-5; 2016.

\bibitem{Buehrer}
Buehrer R.~Michael. Generalized Equations for Spatial Correlation for Low to
  Moderate Angle Spread.  In: Wireless Personal Communications: Bluetooth and
  Other Technologies:101--108Springer US; 2002; Boston, MA.

\bibitem{Mirsad}
Cirkic Mirsad. Efficient MIMO Detection Methods.
\newblock PhD thesisLinkoping UniversityLinkping University, Communication
  Systems, The Institute of Technology2014.

\bibitem{Kobayashi_OSIC}
Kobayashi Ricardo~Tadashi, Abr{\~a}o Taufik. Ordered MMSE---SIC via Sorted QR
  Decomposition in Ill Conditioned Large-scale MIMO Channels.  {\it Telecommun.
  Syst.. }2016;63(2):335--346.

\bibitem{LLL}
Lenstra {A. K.}, Lenstra {H. W.}, Lov{\'a}sz L.. Factoring polynomials with
  rational coefficients.  {\it Mathematische Annalen. }1982;261(4):515--534.

\bibitem{Milford}
Milford D., Sandell M.. Simplified Quantisation in a Reduced-Lattice MIMO
  Decoder.  {\it Communications Letters, IEEE. }2011;15(7):725-727.

\bibitem{Golub}
Golub Gene~H., Loan Charles F.~Van. {\it Matrix Computations}.
\newblock Baltimore, USA: JH Univ. Press; third~ed.1996.

\bibitem{SD_complexity}
Jalden J., Ottersten B.. On the complexity of sphere decoding in digital
  communications.  {\it IEEE Transactions on Signal Processing.
  }2005;53(4):1474-1484.

\bibitem{LLL_Comp}
Ling C., Howgrave-Graham N.. Effective LLL Reduction for Lattice Decoding.  In:
  2007 IEEE International Symposium on Information Theory:196-200; 2007.

\end{thebibliography}
\end{document}